\documentclass[twocolumn,showpacs,preprintnumbers,nofootinbib,prd,superscriptaddress,10pt]{revtex4-1}
\usepackage{graphicx,amssymb,amsmath,amsthm,amsfonts,epsfig}

\usepackage[linktocpage]{hyperref}
\usepackage[usenames,dvipsnames]{color}
\usepackage{epstopdf}
\usepackage{amsmath}
\usepackage{aas_macros}
\usepackage{pifont}
\usepackage{subfig}
\definecolor{darkred}{rgb}{0.5,0,0}
\definecolor{darkgreen}{rgb}{0,0.5,0}
\definecolor{darkblue}{rgb}{0,0,0.5}
\definecolor{prussian}{rgb}{0.0, 0.19, 0.33}
\definecolor{richelectricblue}{rgb}{0.03, 0.57, 0.82}
\definecolor{teal}{rgb}{0.0, 0.5, 0.5}
\definecolor{mediumseagreen}{rgb}{0.24, 0.7, 0.44}
\definecolor{lust}{rgb}{0.9, 0.13, 0.13}
\definecolor{ballblue}{rgb}{0.13, 0.67, 0.8}
\definecolor{darkcyan}{rgb}{0.0, 0.55, 0.55}
\definecolor{mountainmeadow}{rgb}{0.19, 0.73, 0.56}
\definecolor{palecarmine}{rgb}{0.69, 0.25, 0.21}
\definecolor{richcarmine}{rgb}{0.84, 0.0, 0.25}
\definecolor{tangelo}{rgb}{0.98, 0.3, 0.0}
\definecolor{venetian}{rgb}{0.784,0.031,0.082}
\definecolor{bdfrance}{rgb}{0.192,0.549,0.906}

\hypersetup{colorlinks=true, citecolor=blue,
linkcolor=blue, urlcolor=blue}
\usepackage{amsmath,amssymb}
\usepackage{tensor}
\usepackage{mathtools}
\usepackage{amsbsy}
\usepackage{bm}
\usepackage{float}
\newcommand{\be}{\begin{equation}}
\newcommand{\ee}{\end{equation}}
\newcommand{\bear}{\begin{eqnarray}}
\newcommand{\eear}{\end{eqnarray}}

\begin{document}

%\preprint{APS/123-QED}

\title{Prospects of Detecting the Nonlinear Gravitational Wave Memory}% Force line breaks with \\

\author{Aaron D. Johnson}
\affiliation{%
	Department of Physics, University of Arkansas, Fayetteville, Arkansas 72701, USA
}%

\author{Shasvath J. Kapadia}
 \affiliation{%
 	Department of Physics, University of Arkansas, Fayetteville, Arkansas 72701, USA
 }%

 \affiliation{%
Department of Physics, University of Wisconsin -- Milwaukee, CGCA, P.O. Box 413, Milwaukee, Wisconsin 53201, USA
 }%

\author{Andrew Osborne}
\affiliation{%
Department of Physics, University of Arkansas, Fayetteville, Arkansas 72701, USA
}%

\author{Alex Hixon}
\affiliation{%
Department of Physics, University of Arkansas, Fayetteville, Arkansas 72701, USA
}%

\author{Daniel Kennefick}
\affiliation{%
Department of Physics, University of Arkansas, Fayetteville, Arkansas 72701, USA
}%
\affiliation{%
Arkansas Center for Space and Planetary Sciences, University of Arkansas, Fayetteville, Arkansas 72701, USA
}%

\noaffiliation

\date{\today}

\begin{abstract}

In GW150914, approximately $3M_{\odot}$ were radiated away as gravitational waves from the binary black hole system as it merged. The stress energy of the gravitational wave itself causes a nonlinear memory effect in the detectors here on Earth called the Christodoulou memory. We use an approximation that can be applied to numerical relativity waveforms to give an estimate of the displacement magnitude and the profile of the nonlinear memory. We give a signal to noise ratio for a single GW150914-like detection event, and by varying the total mass and distance parameters of the event, we find distances and source masses for which the memory of an optimally oriented GW150914-like event would be detectable in aLIGO and future detectors.

\end{abstract}

% PACS, the Physics and Astronomy Classification Scheme.
\maketitle

%\tableofcontents

%%%%%%%%%%%%%%%%%%%%%%

\section{Introduction}
\label{sec:intro}
Recently reported observations \cite{GW150914, GW151226, LVT151012, GW170104, GW170608, GW170814, GW170817, GWTC1} of gravitational waves in aLIGO consist of black hole binaries and one binary neutron star (BNS) system. Binary systems which lose components are known to produce a memory \cite{Zel, BragGr, BragTh}. Gravitational bremsstrahlung results in a linear memory in a detector far from the source: a permanent displacement between freely falling test masses that grows as the wave passes and persists even after the wave has passed.

Christodoulou (by using the full nonlinear theory of relativity) \cite{Christodoulou} and Blanchet and Damour (by using a post-Minkowskian scheme) \cite{blanDam} independently discovered that the difference in relative position between ideal (freely falling) test masses long before and long after any gravitational wave has passed a detector is nonzero. This memory effect is known as the Christodoulou or nonlinear memory.

\begin{figure}[h]
	\includegraphics[width=80mm]{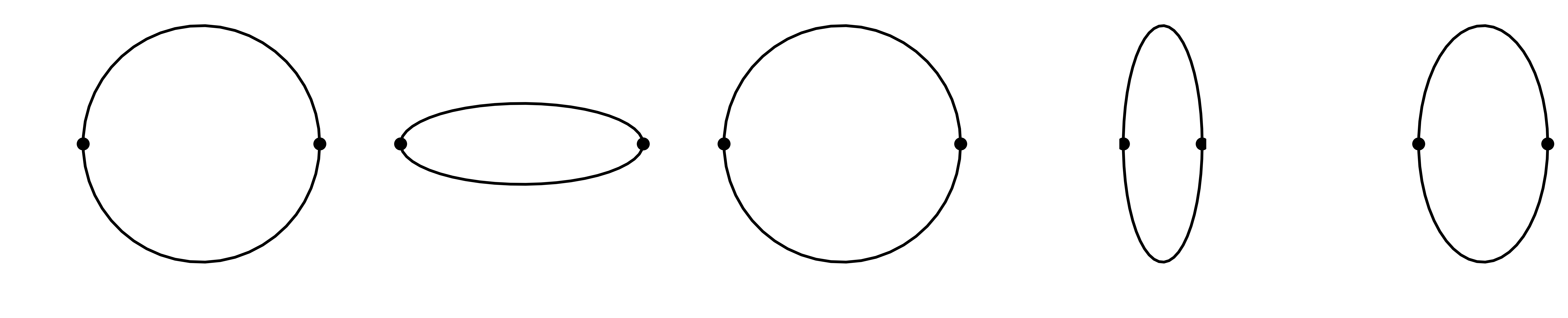}
	\centering
	\caption{Two particles from a ring of particles as a plane, plus polarized gravitational wave passes perpendicular to the page. The nonlinear memory is shown here as a residual plus polarization after the wave has passed.}
	\label{nonlin}
\end{figure}

This is contrary to the standard picture of a gravitational wave that one usually imagines: a ring of particles subject to a plane gravitational wave will oscillate either in a plus or cross polarization pattern and then come back to its original orientation as a ring. In reality, the ring does not return to its original position but is instead left in a residual polarization state as in Figure \ref{nonlin}. Two particles on the ring will either be closer or further apart depending on the sign of the memory. Knowledge of this sign clearly requires knowledge of the polarization state of the oscillatory part of the gravitational wave.

Thorne found that the oscillatory part of the gravitational wave causes the nonlinear memory by considering the wave to be made of gravitons causing a linear memory as they escape from the system \cite{Thorne}. Indeed, a compact binary system loses components in the form of gravitational radiation (about three solar masses in GW150914) thereby causing a nonlinear memory. Since the stress energy of the oscillating gravitational wave as it escapes to infinity causes the nonlinear memory, it can be thought of as the ``wave of the wave'' \cite{WiseWill}.

One may worry that LIGO is insensitive to permanent or DC changes in its arms. How then might we see such an effect? LIGO contains stationary masses held in place, but the detectors will still be able to detect the changing strain caused by the buildup of the memory \cite{Lasky}. Provided the compact binary system is close enough and the inclination angle is optimal, the nonlinear memory could be directly detectable in ground based detectors if enough of the change associated with the memory occurs on a timescale $\tau \approx 1/f_\text{opt}$ where $f_\text{opt}$ is the frequency of the detector's peak sensitivity \cite{Thorne}.

When one of us worked on this in the 1990s, he looked at the detectability of the Christodoulou memory in binary black hole (BBH) systems with masses of at most 10 M$_\odot$ \cite{Kennefick}. Recent detections show most of the BBH systems detected have components with masses above 10 M$_\odot$ \cite{GWTC1}\footnote{Prior to the first detection some physicists believed that the total mass of the black hole binary system would be around 5 - 15 M$_\odot$. Even after the detection, Broadhurst, Diego, and Smoot argue that the sizes of detected BBHs have been exagerated through gravitational lensing \cite{BDS}.}. Now that the oscillatory part of the gravitational wave has been detected, the question has arisen whether the nonlinear memory could be detectable with current sensitivities in ground based detectors \cite{Lasky}.

After GW150914, Lasky, Thrane, Levin, Blackman, and Chen (LTLBC) \cite{Lasky} used Favata's minimal waveform model (MWM) \cite{Favata09a} to find a nonlinear memory waveform. They found a signal to noise ratio (SNR) of 0.42 for an optimally oriented source modeled by the MWM in aLIGO at design sensitivity. A signal with this SNR is not detectable without a clever scheme of adding subthreshold signals as discussed in their paper.

Computation of the nonlinear memory from numerical waveforms has proved difficult, but some calculations have been done for equal mass binary systems \cite{PolReismem}. Extraction of the memory waveform from numerical data to leading order requires two numerical integrations of the $l=2$, $m=0$ part of the Weyl scalar $\psi_4$, the typical output of a numerical simulation. Each integration increases the amplitude of the numerical error until it swamps small, low frequency effects in the signal such as the nonlinear memory \cite{ReisPolint}. Current attempts use Cauchy characteristic matching (see Sec 6.2 in \cite{wini}) to attempt to get more accurate modes containing the memory. However, there is always a piece that is missing from these calculations due to the computation extending only to finite times in the past (we miss the entire inspiral phase before our simulation starts).

While not likely to be detectable by aLIGO, advanced Virgo (AdV), or KAGRA (a Japanese, cryogenic, underground detector which will be operational around 2018 \cite{KAGRA}), one may wonder what the odds of detection are with future detectors. Might we be able to detect the memory as strain sensitivity increases in ground based detectors? Third generation, ground based detectors including the Einstein Telescope (ET) reduce seismic noise by being set up deep underground. The arms of ET are planned to be 10 km long in a triangular geometry with three detectors each comprised of a low frequency, cooled detector and a high frequency detector \cite{ET}. On the same timescale is the Cosmic Explorer (CE) which has an ``L" shape like aLIGO and has 40 km long arms \cite{CE}.

Current ground based gravitational wave detectors hit a wall of seismic noise at about $10$ Hz. Given that the memory is primarily a low frequency effect, perhaps space based detectors sensitive in the decihertz frequency regime could detect the memory. A Japanese, space based detector, DECIGO, is proposed to launch on a timescale similar to LISA \cite{DECIGO}. This gravitational wave detector has three 1000 km arms set up in a triangular pattern and uses differential Fabry-Perot interferometry. However, this timescale includes launching DECIGO pathfinder in 2015, which did not happen. It is sensitive to signals around the decihertz frequency range, filling the frequency gap between the LISA and LIGO detectors. Space based transponder type detectors such as the planned LISA mission are sensitive in the mHz regime and will detect memories of larger binary systems such as extreme mass ratio inspirals (EMRIs) \cite{Favata09a}. Many decades from now, we may see the Big Bang Observer (BBO) launched. BBO consists of smaller LISA type detectors situated in specific ``constellations'' around the sun \cite{BBO}. DECIGO and BBO have strong sensitivity in regions that make them suitable for detecting the memory from a GW150914-like event.  

Pulsar Timing Arrays (PTA) are sensitive to even lower frequencies than space based interferometers and are checked for memory signals from supermassive black hole binaries (SMBH) \cite{PTA}. Among the currently operating PTA groups are the Parkes Pulsar Timing Array (PPTA), the European Pulsar Timing Array (EPTA), and the North American Nanohertz Observatory for Gravitational Waves (NANOGrav). For a recent review of these collaborations, see \cite{PTArev}. The International Pulsar Timing Array group aims to combine the observational data from each group to get even better sensitivity and a greater number of pulsars \cite{IPTA}. In the future the Square Kilometer Array could detect many more pulsars to be used for data analysis and push sensitivity curves even deeper into the noise \cite{SKA}. 

The nonlinear memory is interesting as a purely strong-field gravitational effect. As such, its effects are dependent on the form of Einstein's equations and therefore are useful in theory testing. For example, theories which include scalar fields also contain extra memory modes \cite{mod}. Further, since the rise time of the memory is related to the radii of the compact binary constituents, detection of neutron star binary memory could give independent insight into the equation of state by picking a mass-radius relation and calculating the memory \cite{Kennefick}.

In this paper, we aim to give an approximation of the memory and its profile. We apply Thorne's formula (Equation (3) from  \cite{Thorne}) to a numerical relativity waveform (Section \ref{GW150914}). Using this model results in a calculation that is easy to use and computationally cheap. We calculate the memory for GW150914 (Section \ref{GW150914}). Next, the memory obtained from this calculation can be used to give a signal to noise ratio for a given detector (Section \ref{SNR}). Finally, we vary the mass and distance parameters on a GW150914-like event to find the distance and total source mass for which an event would be detectable in several current and future detectors (Section \ref{massdist}). Throughout this paper we use geometric units ($G=c=1$).

\section{The nonlinear memory of GW150914}
\label{GW150914}

Thorne gives a formula for ``practical computations" of the memory in \cite{Thorne},
\begin{equation}
h_\text{mem}(t) = \frac{2}{r}\int_{-\infty}^{t}dt'\int d\Omega' \frac{d^2E}{dt'd\Omega'}(1+\cos{\theta'})e^{2i\phi'},
\end{equation}
where $r$ is the distance from source to detector, and $d\Omega'$ is the solid angle. By expanding the energy flux in terms of spherical harmonics and integrating out the angular dependence, we can approximate the flux as a prefactor multiplied by the Isaacson stress energy. The full calculation is given in Appendix \ref{CalcMem} and gives an approximation for the memory in optimal orientation,
\begin{equation}
h_\text{max}(t) = \frac{r}{4\pi} \int_{-\infty}^{t} dt' \, \dot{h}^2_\text{+},
\end{equation}
where $h_\text{+}$ is the plus polarization of the oscillatory gravitational waveform from a numerical relativity simulation. The memory amplitude scales depending on inclination \cite{Kennefick} as
\begin{equation}
\Phi(\iota) = \frac{18}{17}\sin^2 \iota \left(1 - \frac{\sin^2 \iota}{18}\right).
\end{equation}
This prefactor shows larger memory effects for edge-on binary systems in contrast to the primary oscillatory wave which is strongest from face-on binaries. In systems that exhibit maximum memory ($\iota = \pi/2$), we will see only half of the maximum $h_\text{+}$ polarization and none of the $h_\times$ polarization from the oscillatory part of the gravitational wave. The memory effect is not present in face-on systems ($\iota = 0$). Using $\Phi (\iota)$, we can find the memory amplitude at any inclination:
\begin{equation}
	h_\text{mem}(t) = \Phi(\iota)\frac{r}{4\pi}\int_{-\infty}^t dt' \, \dot{h}^2_\text{+}.
\end{equation}

\begin{figure}[h]
	\includegraphics[width=80mm]{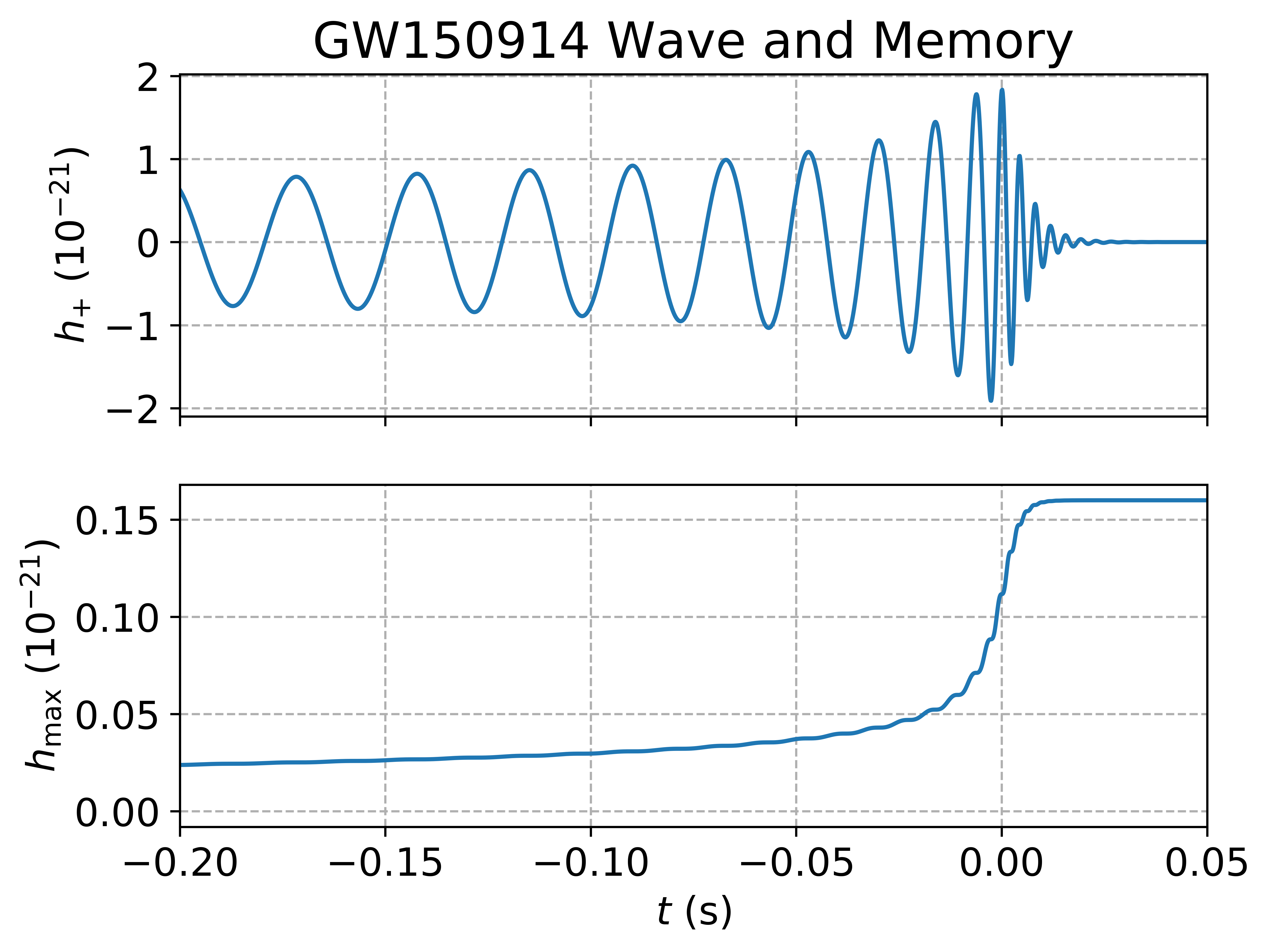}
	\centering
	\caption{Numerical relativity waveform for GW150914 generated with PyCBC is shown in the upper plot. The memory is shown with optimal inclination in the lower plot with maximum amplitude even after the gravitational wave has passed. The ``wiggliness" of the waveform has been discussed in Appendix C of \cite{Favata2012} and is caused by disregarding the average after taking the time integral (see Appendix \ref{CalcMem}).}
	\label{TimeMem}
\end{figure}

By using the method summarized above on the numerical relativity data given by a waveform generated with SEOBNRv4 \cite{SEOBNRv4} using the PyCBC python package \cite{pycbc1, pycbc2} and the LIGO Algorithm Library (LAL), the nonlinear memory can be calculated for GW150914 as shown in Figure \ref{TimeMem}. This waveform uses the averaged parameters given in the spin precessing parameter estimation paper \cite{spinmodel}.

\begin{figure}[h]
	\includegraphics[width=80mm]{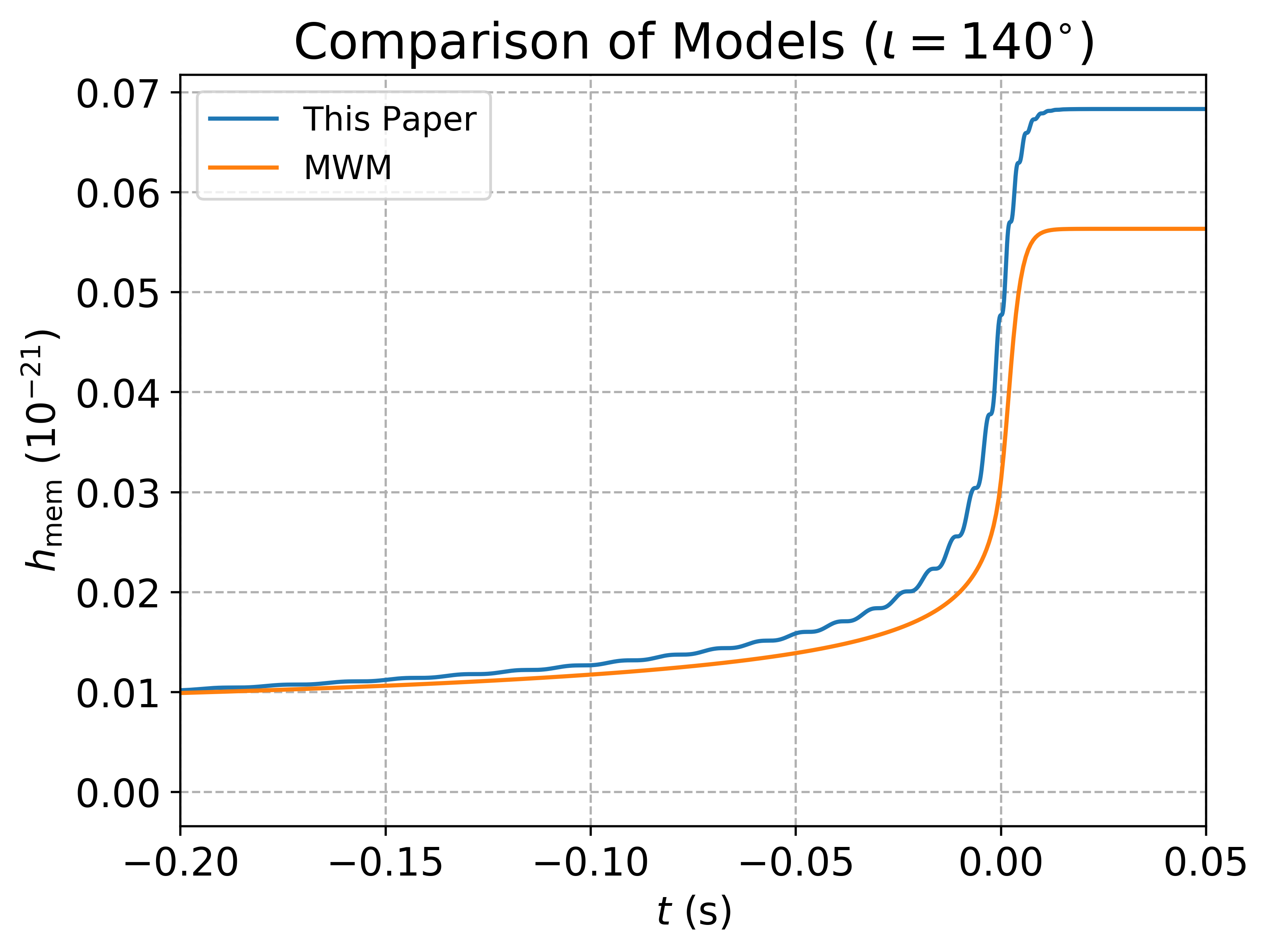}
	\centering
	\caption{Comparison of the MWM in \cite{Lasky} with the approximation given in this paper.}
	\label{MemComp}
\end{figure}

The memory calculated has the same profile as that which LTLBC found with the MWM \cite{Lasky}, as can be seen in Figure \ref{MemComp}. There are two likely reasons for the difference in amplitude. First, our estimate is an overestimate (see Appendix \ref{CalcMem}). Second, the memory we have calculated uses the parameters from \cite{spinmodel} while LTLBC use parameters from \cite{paramest}. The maximum memory has been adjusted to $\iota = 140^{\circ}$ to directly compare between the two models.

\section{Finding the signal to noise ratio}
\label{SNR}
The signal to noise ratio (SNR) denoted $\rho$ can be found by using a Fast Fourier Transform (FFT) routine on the memory waveform. Assuming our template is accurate,
\begin{equation}
\rho^{2} = 4 \int_0^\infty \frac{\left|\tilde{h}(f)\right|^2}{S_n(f)}df,
\end{equation}
where $\tilde{h}(f)$ is the memory signal in the frequency domain, and $S_n(f)$ is the one sided noise power spectral density (PSD) for the detector used to detect the memory signal. These noise curves can be found in the web application listed in the abstract of \cite{GWPlot} with the exception of NANOGrav, which may be found in the 11 year data set \cite{NANOG}. Outside of the frequency range for which any gravitational wave detector noise curve is given, we take the SNR to be zero. For the primary oscillatory wave an SNR of 5 - 8 is sufficient to be confident in detection. Since the memory will accompany the oscillatory part of the wave, papers that have considered memory detection have claimed that an SNR of 3 - 5 would be sufficient to be confident in detection \cite{Lasky} \cite{Kennefick}\footnote{What value of SNR gravitational wave scientists will deem acceptable in the future is largely a sociological issue. An example of the tightening of experimental standards for particle physics in the past can be found in \cite{Franklin}.}. Following these other papers, we plot SNR values of 3 and 5 in all figures. In the event that we are looking for memory by itself, we also plot an SNR value of 8.

\subsection{SNR for GW150914-like events in current and future detectors}
\begin{table}[h]
	\centering
	\begin{tabular}{|r|c||r|c|}
		\hline
		Detector        & SNR & Detector        & SNR \\ \hline
		AdV				& 0.238  &  eLISA    & 0.025    \\ \hline
		aLIGO           & 0.450  &  LISA     & 0.214     \\ \hline
		KAGRA           & 0.243  &    &   \\ \hline
		ET              & 9.726  &  DECIGO   & 96.53  \\ \hline
		CE              & 27.73  &  BBO      & 177.2  \\ \hline		
	\end{tabular}
	\caption{Current and future detectors' SNR for optimally oriented GW150914-like memory events}
	\label{ground}
\end{table}
As has been discussed in the paper by LTLBC \cite{Lasky}, the memory is not likely to be detectable in aLIGO without adding several detections together. We agree with their results with a calculated SNR of 0.45 for an optimally oriented memory source (compared with an SNR of 0.42 found in \cite{Lasky}). Therefore, we need a closer event to detect the memory with aLIGO's current sensitivity. 

Using the distance that gives $\rho = $ 3 (5) and the median event rate from \cite{events} of 17 $ \text{Gpc}^{-3}\text{yr}^{-1}$ events like GW150914, we can find how long we can expect to wait for a detectable GW150914-like memory event in current ground based detectors. A detectable event would occur at 65 (39) Mpc. With current event rates, we expect this to happen once in about 51 (237) years. However, as can be seen in Table \ref{ground}, both third generation ground based detectors will be able to detect the nonlinear memory effect from optimally oriented GW150914-like events.

From Table \ref{ground}, it is apparent that eLISA and LISA will do no better than ground detectors in detecting an event like GW150914. Both space based decihertz detectors, DECIGO and BBO, will certainly be able to see the memory. This is probably due to the frequency band being optimal for this signal. Additionally, these detectors have impressive projected sensitivities. Even with sensitivity reduced by a factor of 10, DECIGO would still see the memory from a GW150914-like event. Neither decihertz detector has a production timeline set, but DECIGO has an optimistic launch date as early as 2027 \cite{DECIGO}. This launch date expected DECIGO pathfinder to launch in mid 2015 as a precursor to DECIGO. Given that this did not take place, the timeline should be adjusted accordingly.

\section{GW150914-like events with varying mass and distance}
\label{massdist}
Since current detectors have little chance of detecting any single GW150914-like event memory, we now vary parameters on the event to find what luminosity distance $d_L$ and total detector frame mass $M$ would be needed for a given detector to achieve a detection\footnote{Here we use the standard $\Lambda$CDM cosmology \cite{Planck15}. This results in a frequency shift which is equivalent to a mass change between the source and detector frames given by $m_{\text{detector}} = (1 + z) m_{\text{source}}$ where $z$ is the redshift \cite{GWTC1}.}. All points on the grid use the same template with varied mass and distance yielding a similar system in a different frequency regime or reducing the SNR as distance increases. Figures \ref{grounddet1}, \ref{grounddet2}, \ref{spacedet1}, \ref{spacedet2}, \ref{PTAdet1}, \ref{PTAdet2} show the SNR for a given $(d_L, M)$. In the cases where it is relevant, the plots show a nearby plot extending to 3000 Mpc in distance on the left, and a plot that extends to the edge of the $\rho=3$ contour or up to a luminosity distance of 30 Gpc on the right. There are contours for $\rho=3$, $\rho=5$, and $\rho=8$ on these plots to show different standards for detection in each detector. Reported events are shown as marks at their estimated $(d_L, M)$ parameters \cite{GWTC1}.

However, GW170817 is a neutron star binary while the SNR values are given based on a binary black hole system. Therefore, we expect the actual memory SNR to be less than projected on the plot and moved further to the right. Yang and Martynov looked at the detectability of binary neutron star mergers with four different equations of state \cite{YangMarty}. They found that two 1.325 $M_\odot$ neutron stars at a distance of 50 Mpc in aLIGO produce an SNR of about $\rho = 0.1$ and an SNR of about $\rho=10$ in CE. These values are consistent with values given in \cite{GWTC1} for GW170817. At the same values, we find $\rho = 0.483$ for aLIGO and $\rho = 18.479$ for CE. We assume that the memory amplitude is off by the same multiplicative factor at all frequencies. Here we pick the worse of the two factors of 4.83. At the median distance and mass, $(d_L,M) = (40, 2.8)$, given in \cite{GWTC1} for GW170817, we find an SNR of around $\rho = 4.772$ in DECIGO and $\rho = 17.68$ in BBO. These results deserve further study with different equations of state.

All SNR values here are calculated assuming a comparable mass binary system. As the frequency band gets lower, the mass ratio of the two black holes may become more extreme. However, for higher mass ratios, the memory is less prominent \cite{Favata09a}.

Figure \ref{grounddet1} shows current and near future detectors unable to detect the nonlinear memory further away than 250 Mpc. Given that the majority of black hole related events have occurred further away than that, it seems unlikely that we will see the memory with current ground based detectors. Outlook for future ground based detection of the memory is positive. Both proposed third generation detectors improve sensitivity and visibility distance significantly.

Even if we don't see the nonlinear memory from current ground based detectors, it should be visible in a different regime with LISA. Supermassive black hole mergers with total mass on the order of $10^7 M_{\odot}$ give a promising source for memory detections as can be seen in Figure \ref{spacedet1}. DECIGO and BBO will be able to see optimally oriented memory from all of the recently reported sources. These detectors also open the exciting possibility of detecting neutron star binary gravitational nonlinear memory.

Pulsar timing arrays NANOGrav, EPTA, and IPTA have not reported detection yet. There is thought to be an upper bound of supermassive black hole mass at around $10^{10} M_{\odot}$ \cite{King}. The fact that no memory has been seen in such detectors supports this as shown in Figure \ref{PTAdet} where we can see that comparable mass binary systems with this mass would produce a memory which should be visible to current pulsar timing arrays.

\section{Conclusion}
Gravitational wave astronomy is in its infancy. Now that the primary oscillatory wave has been detected, we look toward nonlinear parts of the gravitational wave. Current ground based detectors (see Figure \ref{grounddet1}) are unlikely to see the nonlinear memory without a clever stacking scheme as in \cite{Lasky} and \cite{YangMarty}. The outlook for future detectors is positive: third generation ground based detectors (see Figure \ref{grounddet2}), space based detectors (see Figures \ref{spacedet1}, \ref{spacedet2}), and pulsar timing arrays (see Figures \ref{PTAdet1}, \ref{PTAdet2}) can all detect the nonlinear memory in different frequency regimes. DECIGO and BBO in particular yield interesting prospects for detecting the nonlinear memory from a neutron star binary system. This should be considered further with varying equations of state similar to what has been done with CE in \cite{YangMarty}. Given that the rise time of the memory could be related to the radius in some way \cite{Kennefick}, the memory yields an independent method of constraining the neutron star equation of state. Future funding applications for newer, better gravitational wave detectors should include detecting the nonlinear memory and its applications as a science goal.

\section{Acknowledgments}
We thank Kip Thorne for suggesting this project to us, and Stephen Hawking, Malcolm Perry, and Andy Strominger for their query to Kip regarding the memory's detectability. We thank Lasky, Thrane, Levin, Blackman, and Chen for kindly providing their data for straightforward comparison. We would also like to thank Kostas Glampedakis, Scott Hughes, Mark Hannam, Mark Scheel, Xavier Siemens, Jolien Creighton, and Alan Wiseman for useful discussion. AJ and DK would like to thank University College Cork for their hospitality during the preparation of this manuscript. AJ worked as a Sturgis International Fellow funded by the Roy and Christine Sturgis Educational Trust while abroad.

This work has made use of the LIGO Algorithm Library (LAL) and the PyCBC Python package.
\appendix

\section{Calculation of the nonlinear memory}
\label{CalcMem}
Following the derivations for the nonlinear memory in \cite{WiseWill, favata2010}, we begin with the relaxed Einstein field equation in the harmonic gauge,
\begin{equation}
\label{refe}
\Box \bar{h}^{\mu\nu} = -16\pi\tau^{\mu\nu}.
\end{equation}
where $\tau^{\mu\nu}$ contains the stress-energy $T^{\mu\nu}$, the Landau-Lifshitz pseudotensor $t^{\mu\nu}_{LL}$, and some pieces of $\mathcal{O}(h^2)$. Assuming a flat background spacetime and ignoring the other pieces of $\tau^{\mu\nu}$, we focus on this part that sources the memory,
\begin{equation}
T^{gw}_{jk} = \frac{1}{r^2}\frac{d^2E}{dt'd\Omega'}\xi_j \xi_k,
\end{equation}
where $\xi^j$ is a unit vector pointing from the source to solid angle $d\Omega'$ and $d^2E/dt'd\Omega'$ is the gravitational wave flux. Solving equation (\ref{refe}) by using the retarded Green's function yields the standard result
\begin{equation}
	\bar{h}^{jk} = 4 \int \frac{T^{jk}_\text{gw}(t' - |\mathbf{x} - \mathbf{x'}|, \mathbf{x'})}{|\mathbf{x} - \mathbf{x'}|}.
\end{equation}
Next, we use a method of direct integration of the relaxed Einstein equations (DIRE) \cite{PoissonWill}. This method changes the coordinates from Cartesian to spherical and the radial coordinate $r'$ to retarded time $u'=t' - r'$. We then integrate with respect to retarded time in the wave zone. This process yields
\begin{equation}
	\bar{h}_{jk} = \int_{-\infty}^{u}du' \int \frac{d^2 E}{d\Omega'dt'}\frac{\xi_j\xi_k}{t - u' - \mathbf{x}\cdot \xi'}d\Omega'
\end{equation}
where $u = t - r$ and $\xi' = \mathbf{x'}/r'$. This is Equation (4) of \cite{WiseWill}. We now specialize to a gravitational wave burst passing a detector at fixed $r$. Using the limiting procedure in \cite{WiseWill} and transforming to transverse traceless gauge,
\begin{equation}
h_{jk}^\text{TT} = \frac{4}{r}\int_{-\infty}^{t}dt' \int \frac{d^2E}{dt'd\Omega'}\left(\frac{\xi_j\xi_k}{1-\cos\theta'}\right)^\text{TT}d\Omega'.
\end{equation} 
From here Thorne suggested a system of coordinates for ``practical computations" and found a formula for the nonlinear memory. Equation (3) from \cite{Thorne} modified by replacing $dE/d\Omega'$ with $\int_{-\infty}^{t}(d^2E/ dt' d\Omega')dt'$ as in \cite{Kennefick} is
\begin{equation}
\label{Thorne}
h_{\text{mem}} = \frac{2}{r}\int_{-\infty}^t dt'\int d\Omega '\frac{d^2E}{dt'd\Omega '}\left(1 + \cos \theta '\right)e^{2i\phi '}
\end{equation}
where $r$ is the distance from the source to the detector, $t$ is some time after the wave has passed, and $\Omega$ is the solid angle. Notice that we are integrating over the entire history of the wave until now, hence the name ``memory." An expansion in terms of the spin-weighted spherical harmonics,
	\begin{equation}
	\label{fluxdecomp}
	\frac{d^2E}{d\Omega' dt'} = \frac{r^2}{16\pi}\sum_{l,l',m,m'} \left< \dot{h}_{lm}\dot{h}^{*}_{l'm'} \right>\,  _{-2}Y_{lm}\, _{-2}Y_{l'm'}^*
	\end{equation}
where the brackets, $\left<\right>$, denote a time average over several wavelengths of the wave. This allows one to separate the angular piece from the temporal piece. Here the $_{-2}Y_{lm}$ are the spin-weighted spherical harmonics, and the $*$ denotes complex conjugation. Using Equation (\ref{fluxdecomp}) in Equation (\ref{Thorne}), we find the selection rule $m + m' = -2$. After performing the angular integration,
\begin{widetext}
	\begin{equation}
	h_\text{mem}(t) = \frac{r}{8\pi}\int_{-\infty}^{t}dt'\left(\sqrt{\frac{2}{3}}\left< \dot{h}_{20}\dot{h}_{22}^* \right> + \sqrt{\frac{1}{6}}\left< \dot{h}_{2-2}\dot{h}_{20}^* \right> + \frac{2}{3}\left< \dot{h}_{2-1}\dot{h}_{21}^* \right>\right).
	\end{equation}
\end{widetext}
Because of the issues with directly calculating $h_{20}$ modes of the gravitational wave discussed in Section \ref{sec:intro}, we aim for an approximation. We change all smaller modes to the dominant $h_{22}$ piece of the waveform. What we would like to use here is the $(l,m) = (2,2)$ mode that exists at the source. Instead, we are given the mode in the detector's frame. Then,
\begin{equation}
h_\text{mem}(t) = \frac{r}{4\pi}\int_{-\infty}^{t}dt' \left<\dot{h}_{22}^2\right>.
\end{equation}
This equation can be compared to Equation (21) in \cite{favata2010} from which we find the prefactor off by a factor of about 2 (after the inclination has been put in terms of $\sin^2\iota$). The origin of this factor is unknown, but it does not significantly alter the results in a side-by-side comparison. For a direct approximation from the gravitational waves given by a numerical relativity waveform, we can replace the dominant mode with the Isaacson stress-energy,
\begin{equation}
h_\text{mem}(t) = \frac{r}{4\pi}\int_{-\infty}^{t}dt' \left<\dot{h}_+^2 + \dot{h}_{\times}^2\right>.
\end{equation}
By doing so, we're adding in higher order terms to connect our approximation to the waveform given by a numerical simulation. Here we focus our attention to linearly polarized waves at $\iota = \pi/2$. At this inclination, the oscillatory wave only consists of half the amplitude of the $h_{+}$ polarization. The memory is at a maximum, but this amplitude will be based on the maximum amplitude of the oscillatory part of the wave. So as a kludgy model, we use
\begin{equation}
h_\text{max}(t) = \frac{r}{4\pi}\int_{-\infty}^{t}dt' \dot{h}_+^2
\end{equation}
where we have dropped the average since the time integral is effectively taking the average by integrating over the entire history of the wave. This is the cause of the ``wiggliness'' seen in Figure \ref{TimeMem} as discussed in \cite{Favata2012}.

\section{A note on the form of the memory and changing parameters}
\begin{figure}[t]
	\begin{center}
		\Large Current and Near Future Ground Based Detectors
	\end{center}
	\includegraphics[width=85mm]{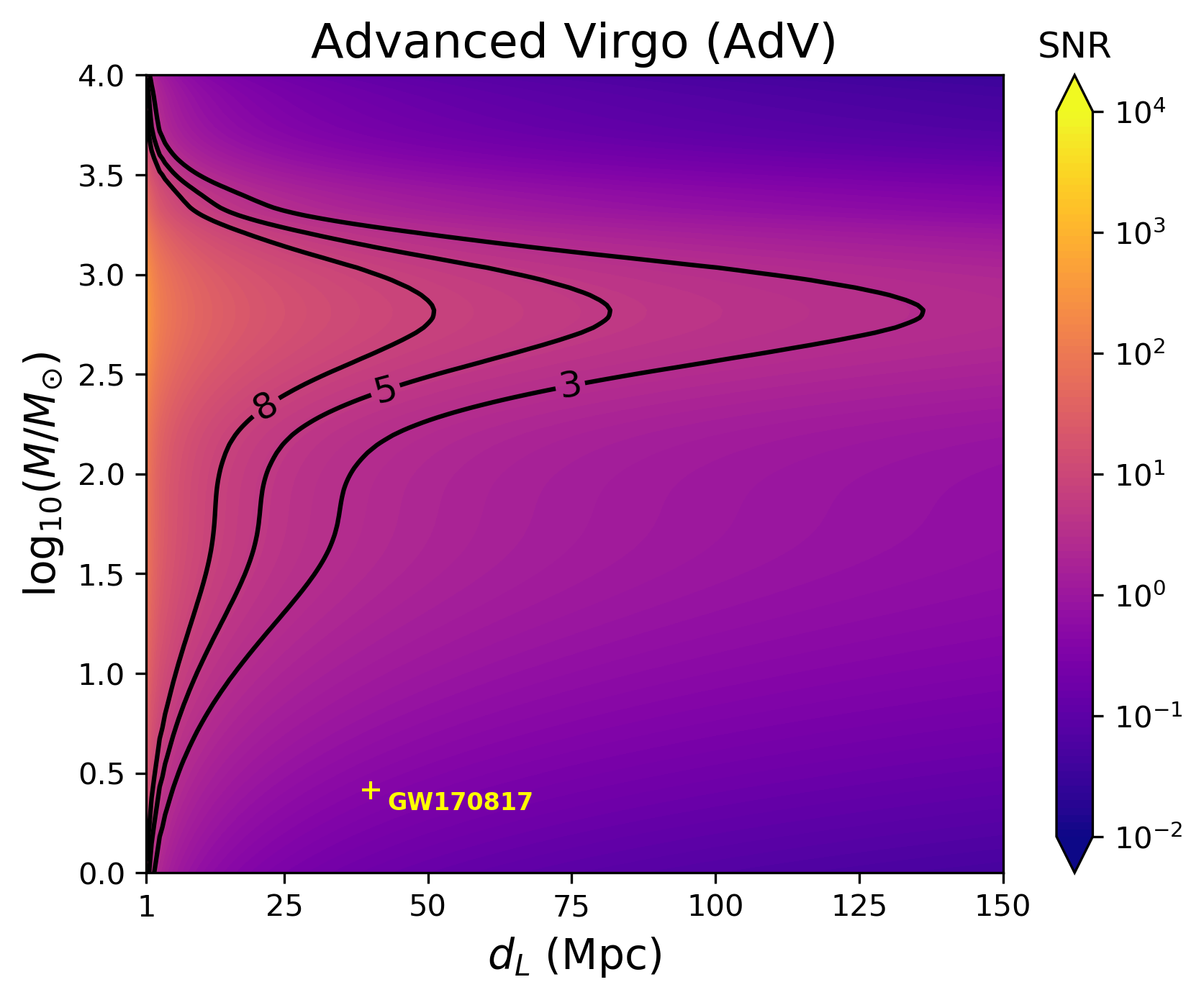}\\
	\includegraphics[width=85mm]{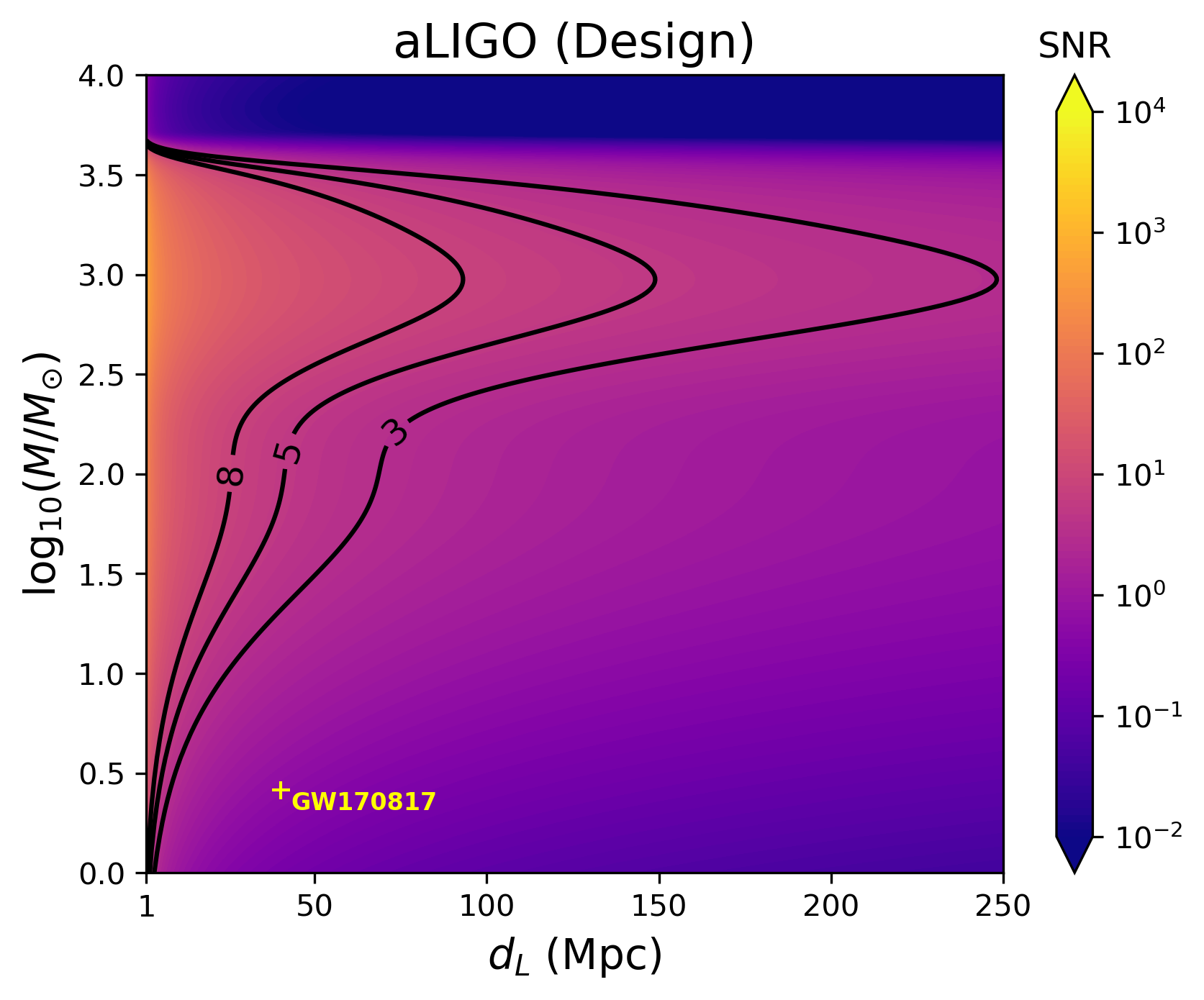} \\
	\includegraphics[width=85mm]{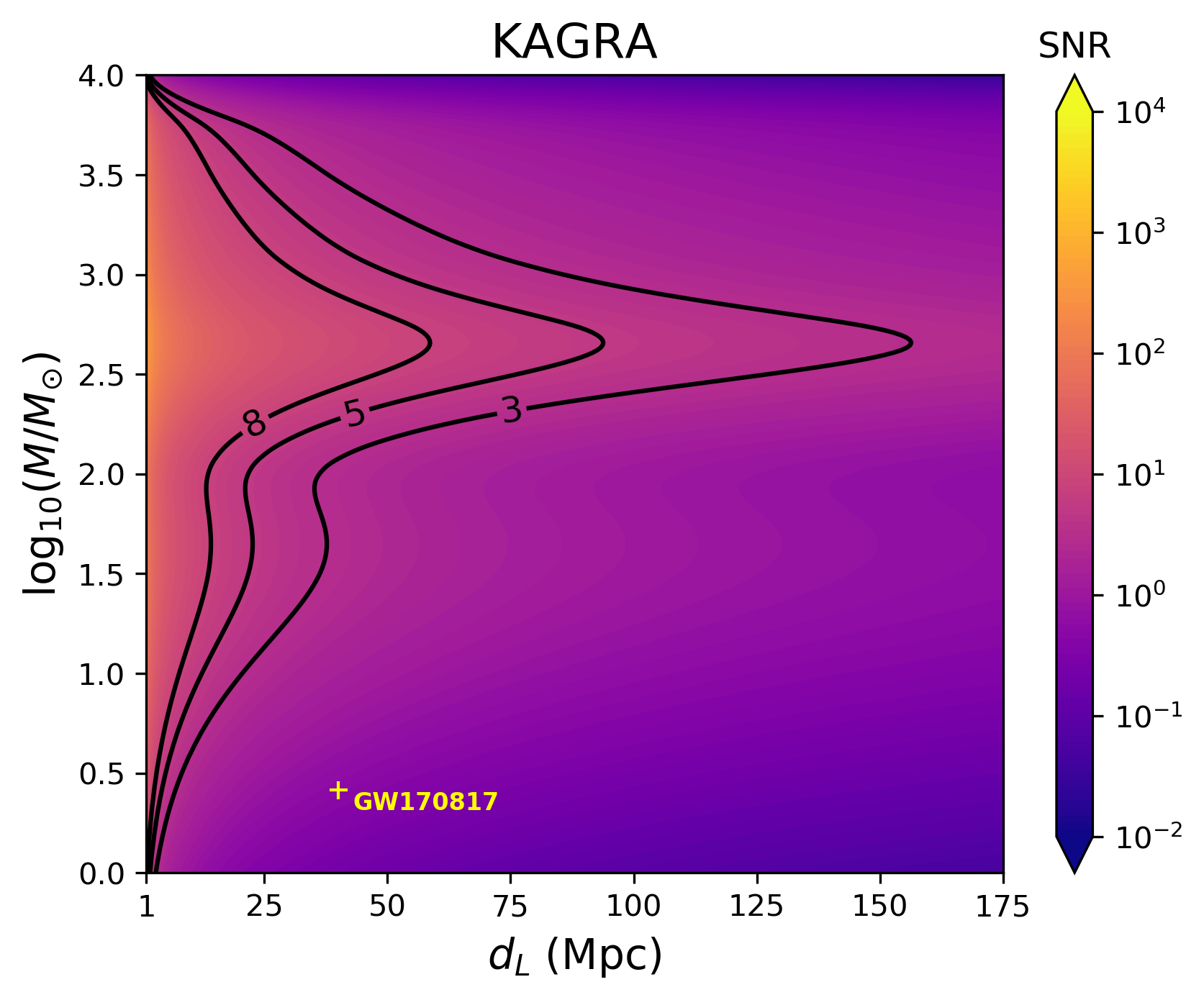}
	\caption{Luminosity distance ($d_L$), total mass (M) parameter space for current and near future ground based detectors. SNR values are shown on a logarithmic color scale for a given $(d_L, M)$ pair. Contours show SNR values of 3, 5, and 8. Events are marked with a ``+" to indicate where they fall on the SNR scale as a BBH, GW150914-like system at optimal orientation. Current and near future ground based detectors (2nd generation detectors) are not able to see memory from any released detection.}
	\label{grounddet1}
\end{figure}

\begin{figure}[t]
	\begin{center}
		\Large Future Ground Based Detectors
	\end{center}
	\begin{tabular}{c | c}
	\textbf{\large Near} & \textbf{\large Far} \\
	\includegraphics[width=85mm]{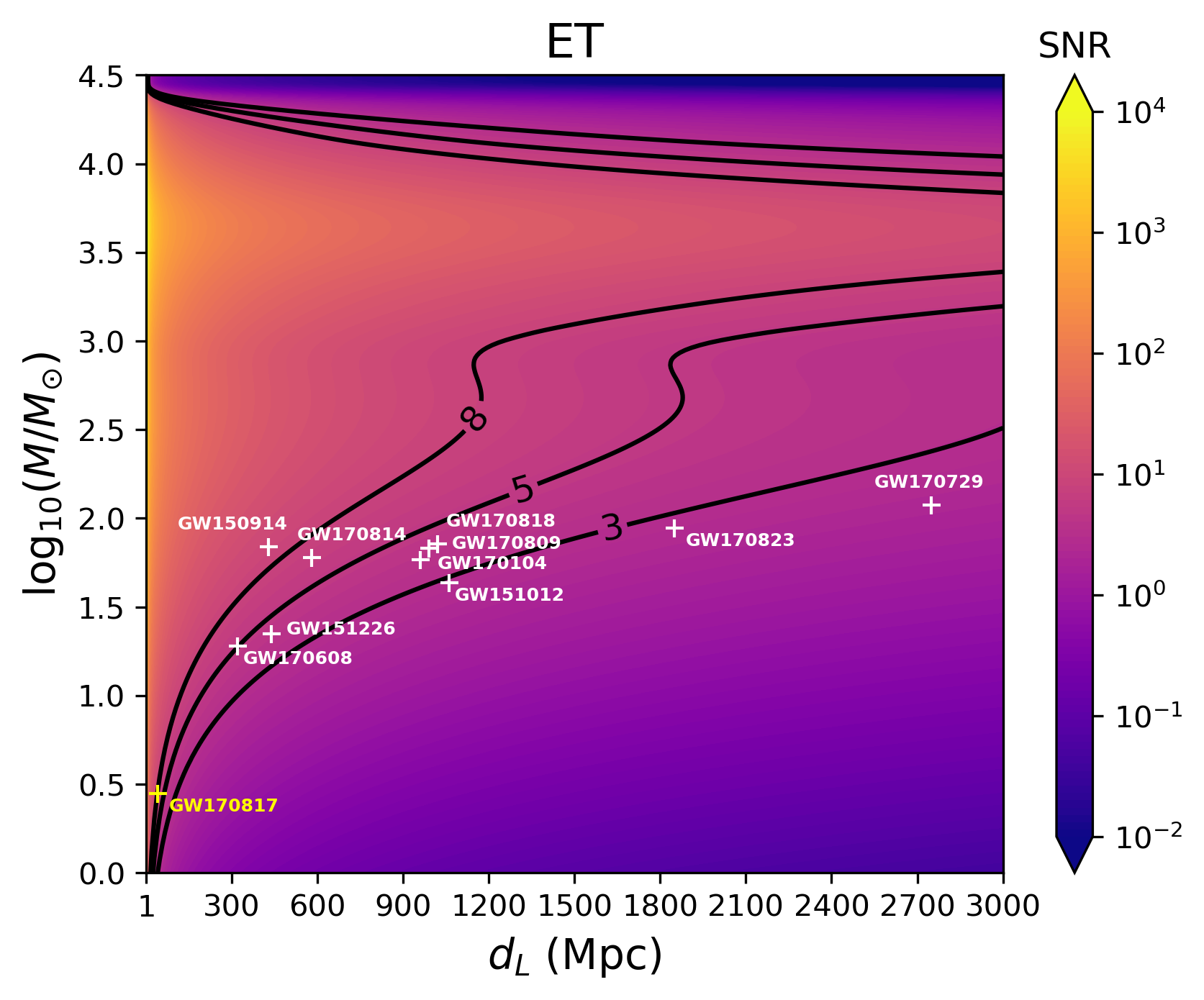} & \includegraphics[width=85mm]{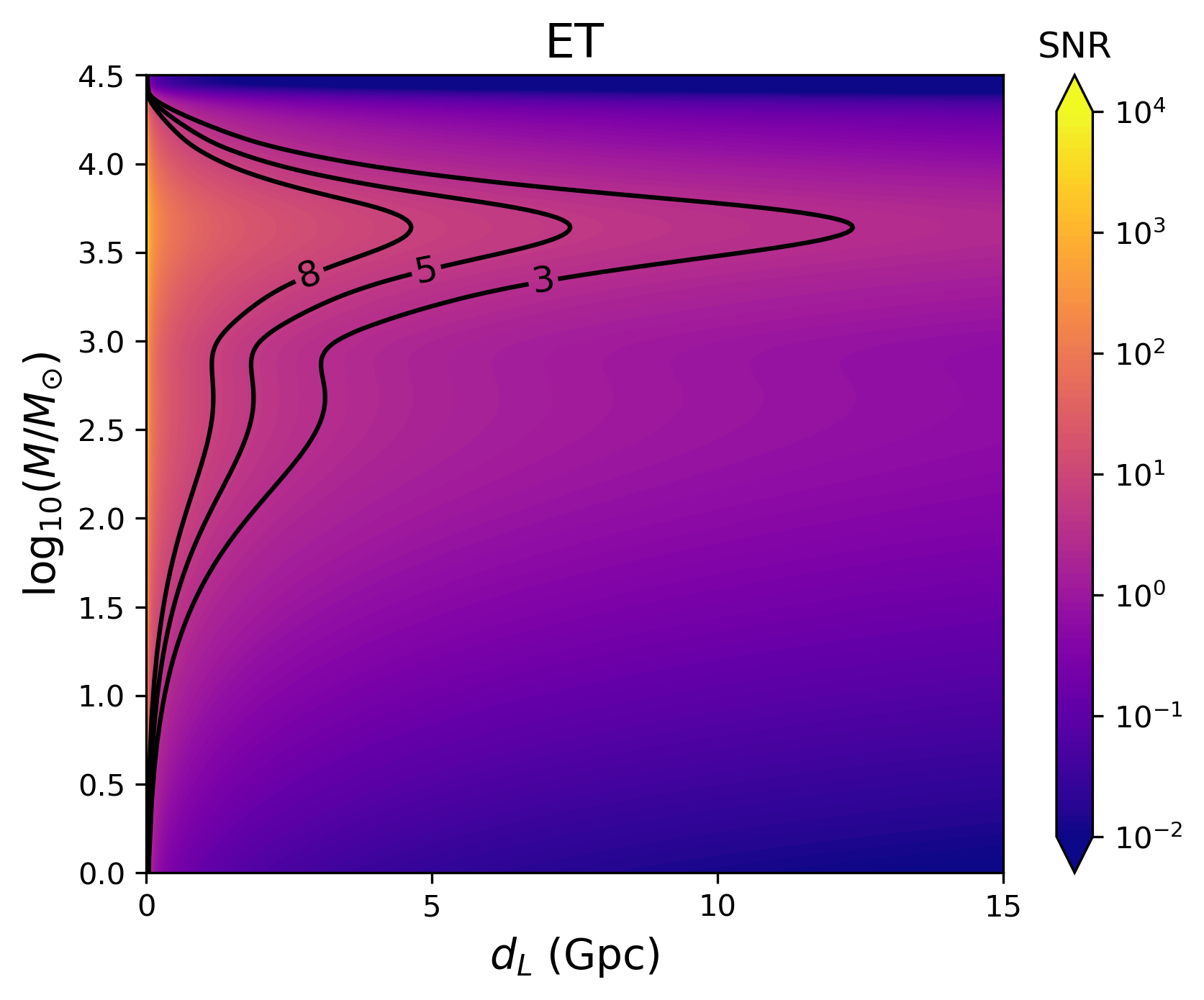} \\
	\includegraphics[width=85mm]{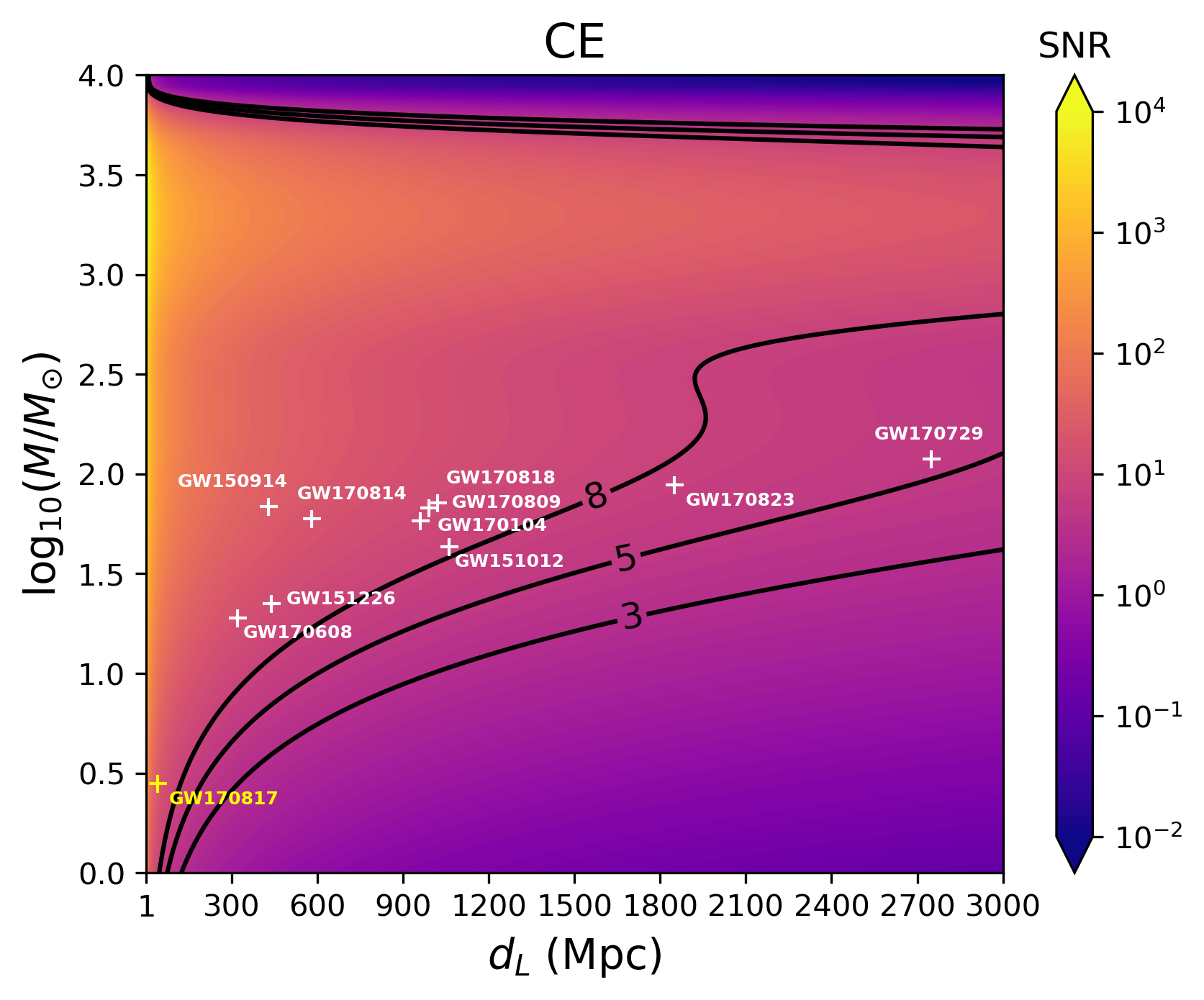} & \includegraphics[width=85mm]{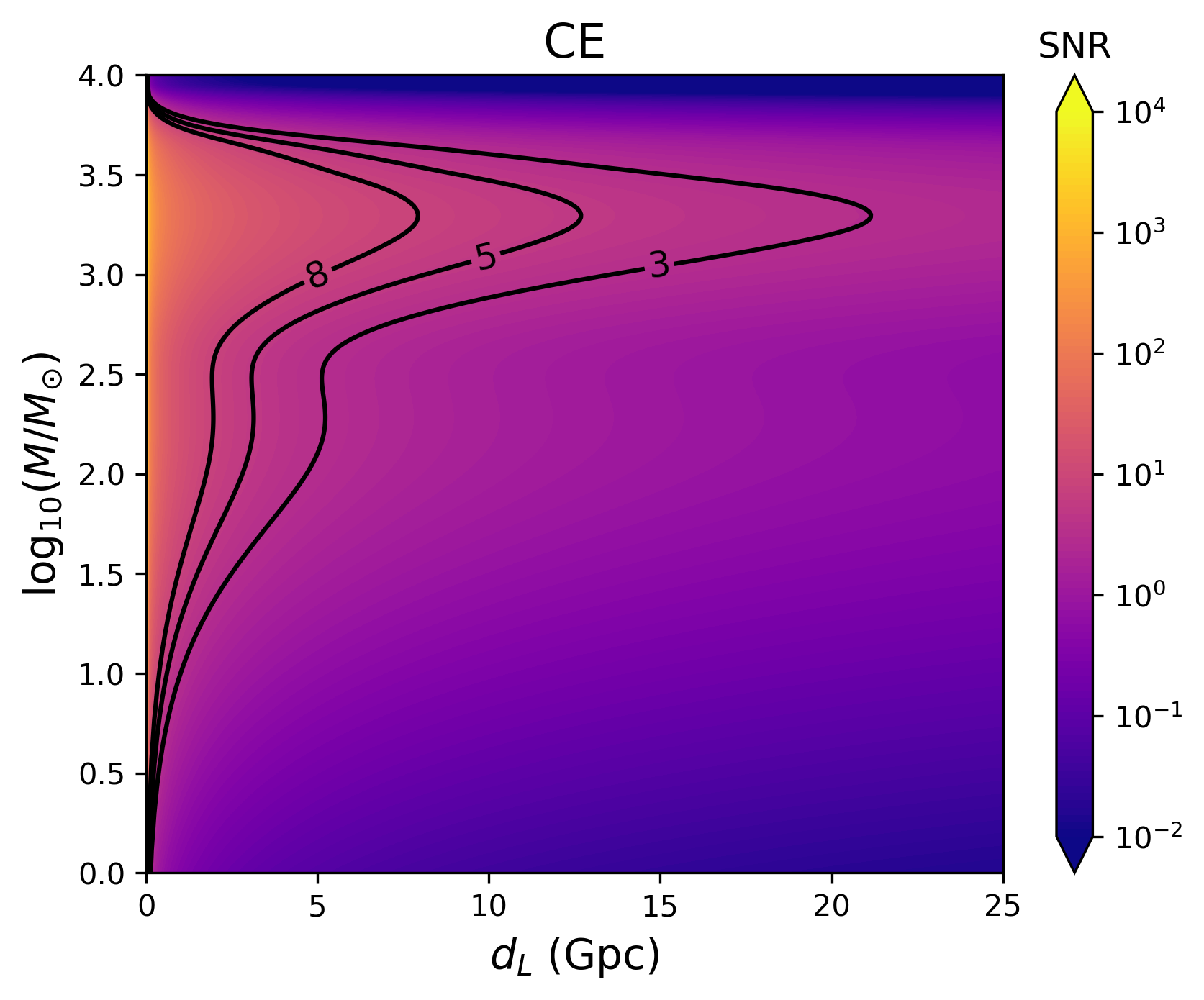}
	\end{tabular}
	\caption{Luminosity distance ($d_L$), total mass (M) parameter space for current and near future ground based detectors. SNR values are shown on a logarithmic color scale for a given $(d_L, M)$ pair. Contours show SNR values of 3, 5, and 8. Events are marked with a ``+" to indicate where they fall on the SNR scale as a BBH, GW150914-like system at optimal orientation. Plots in the left column show SNR values out to 3000 Mpc, and plots on the right column show the edge of the $\rho = 3$ contour. Future ground based detectors (3rd generation detectors) offer significantly improved sensitivity and are able to see memory from several released detections. Although GW170817 is a BNS, Yang and Martynov \cite{YangMarty} show that for various equations of state CE gives an SNR of about 10 and should be detectable.}
	\label{grounddet2}
\end{figure}

\begin{figure} [t]
	\begin{center}
		\Large Space Based mHz Detectors
	\end{center}
	\begin{tabular}{c | c}
	\textbf{\large Near} & \textbf{\large Far} \\
		\includegraphics[width=85mm]{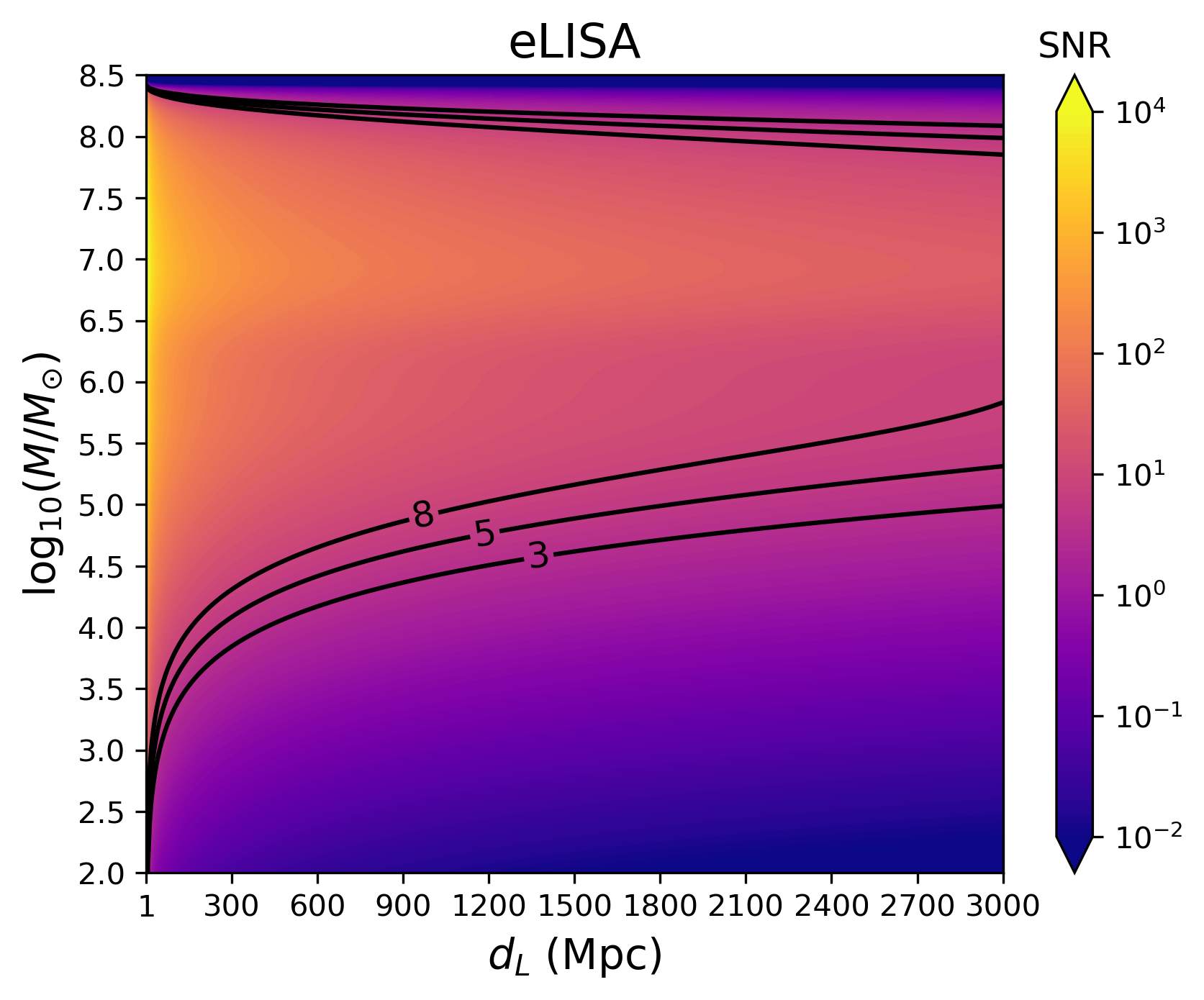} & \includegraphics[width=85mm]{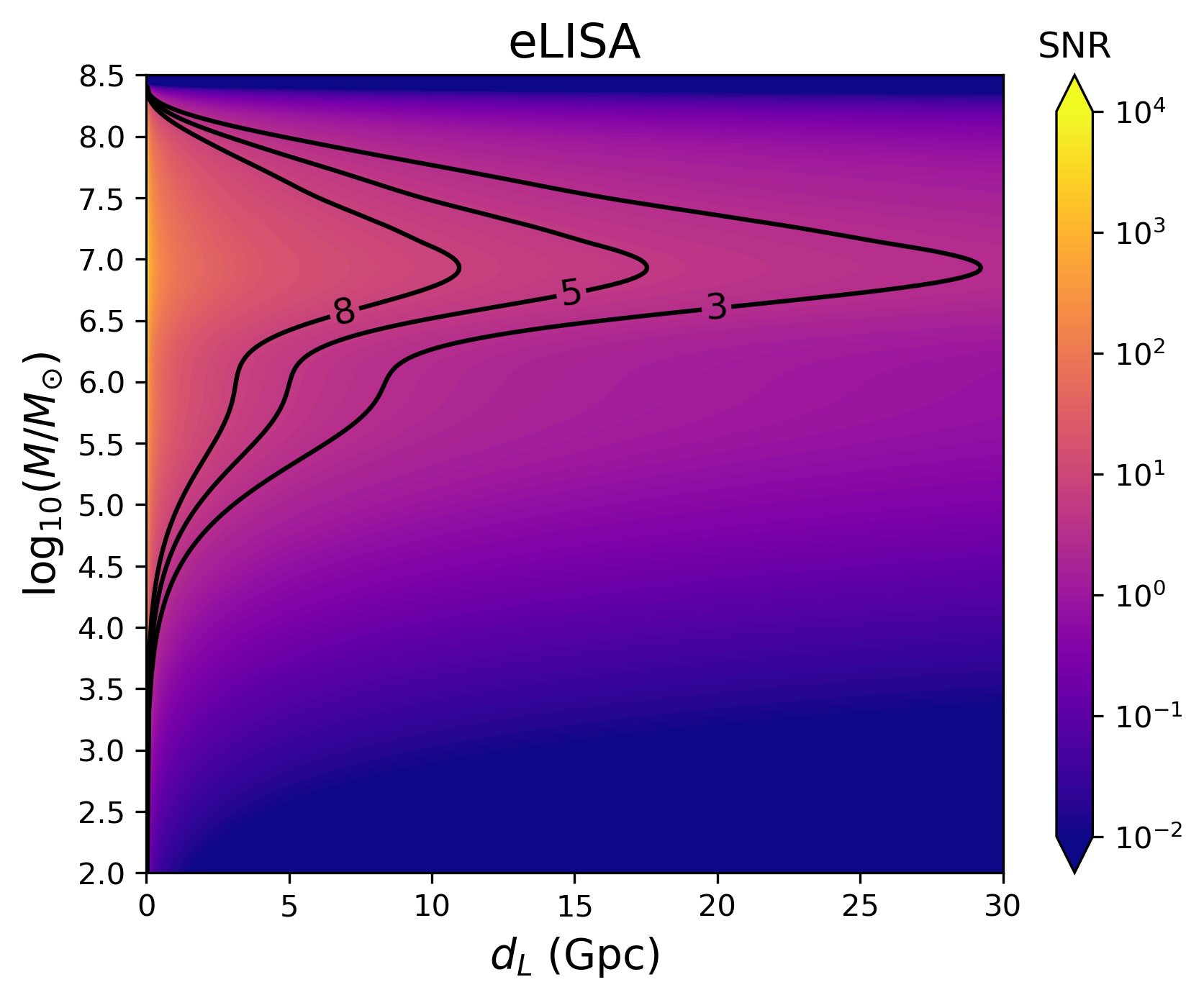} \\
		\includegraphics[width=85mm]{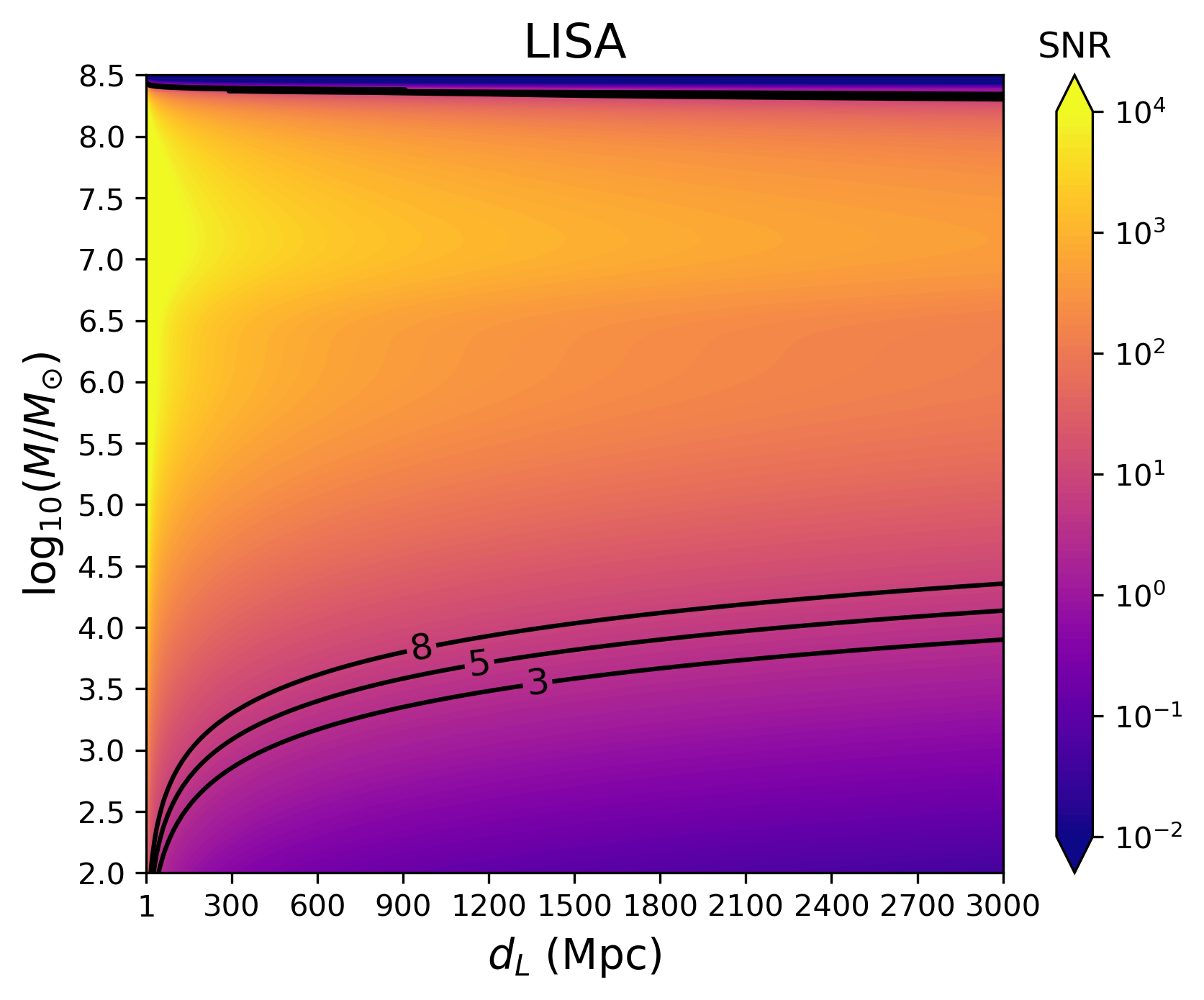} & \includegraphics[width=85mm]{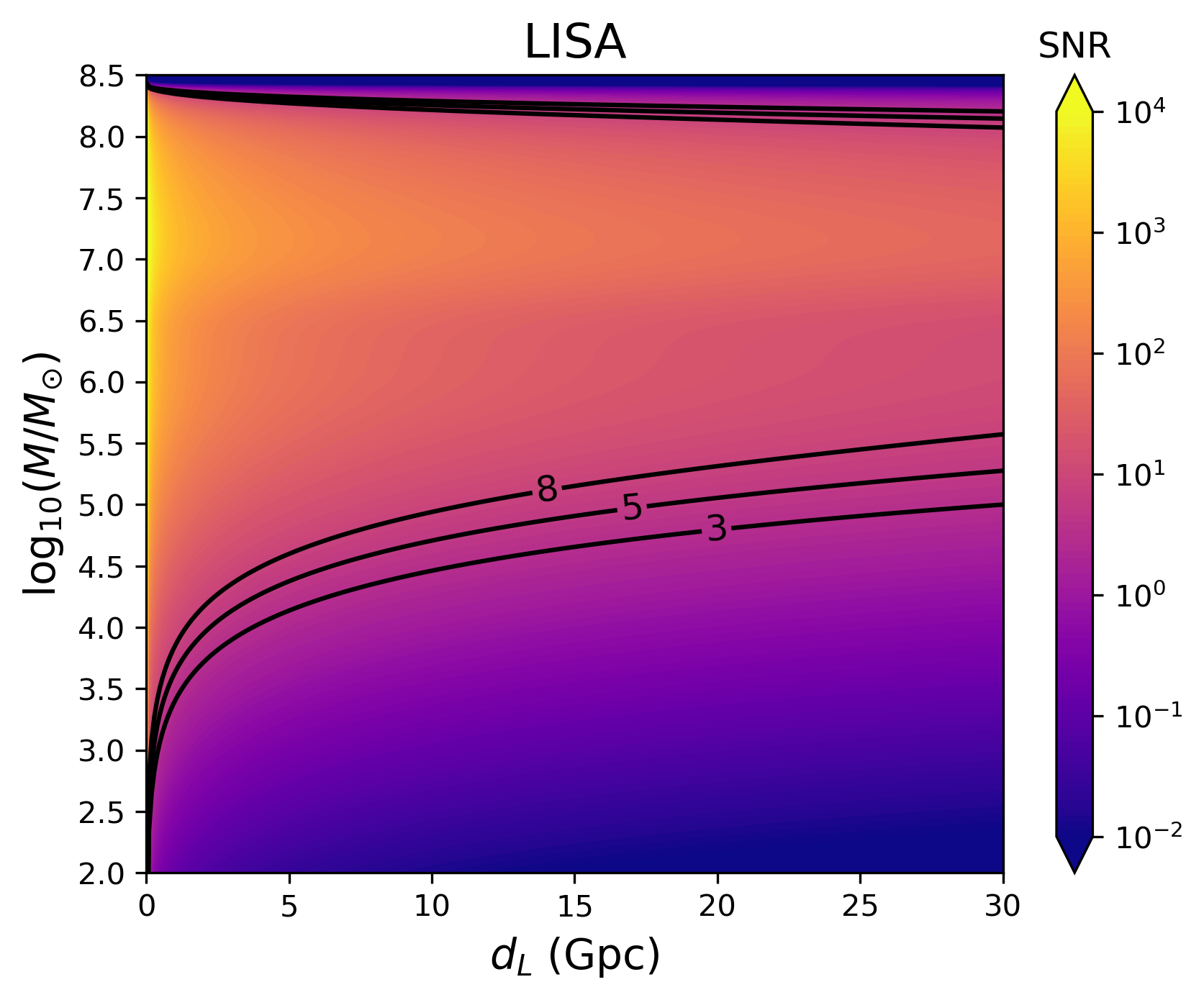}
	\end{tabular}
	\caption{Luminosity distance ($d_L$), total mass (M) parameter space for current and near future ground based detectors. SNR values are shown on a logarithmic color scale for a given $(d_L, M)$ pair. Contours show SNR values of 3, 5, and 8. Events are marked with a ``+" to indicate where they fall on the SNR scale as a BBH, GW150914-like system at optimal orientation. Plots in the left column show SNR values out to 3000 Mpc, and plots on the right column show the edge of the $\rho = 3$ contour or out to 30 Gpc. Space based detectors in the mHz frequency regime will be able to see SMBH binary memory with masses significantly higher than those that are visible in the ground based detectors.}
	\label{spacedet1}
\end{figure}

\begin{figure} [t]
	\begin{center}
		\Large Space Based dHz Detectors
	\end{center}
	\begin{tabular}{c | c}
	\textbf{\large Near} & \textbf{\large Far} \\
		\includegraphics[width=85mm]{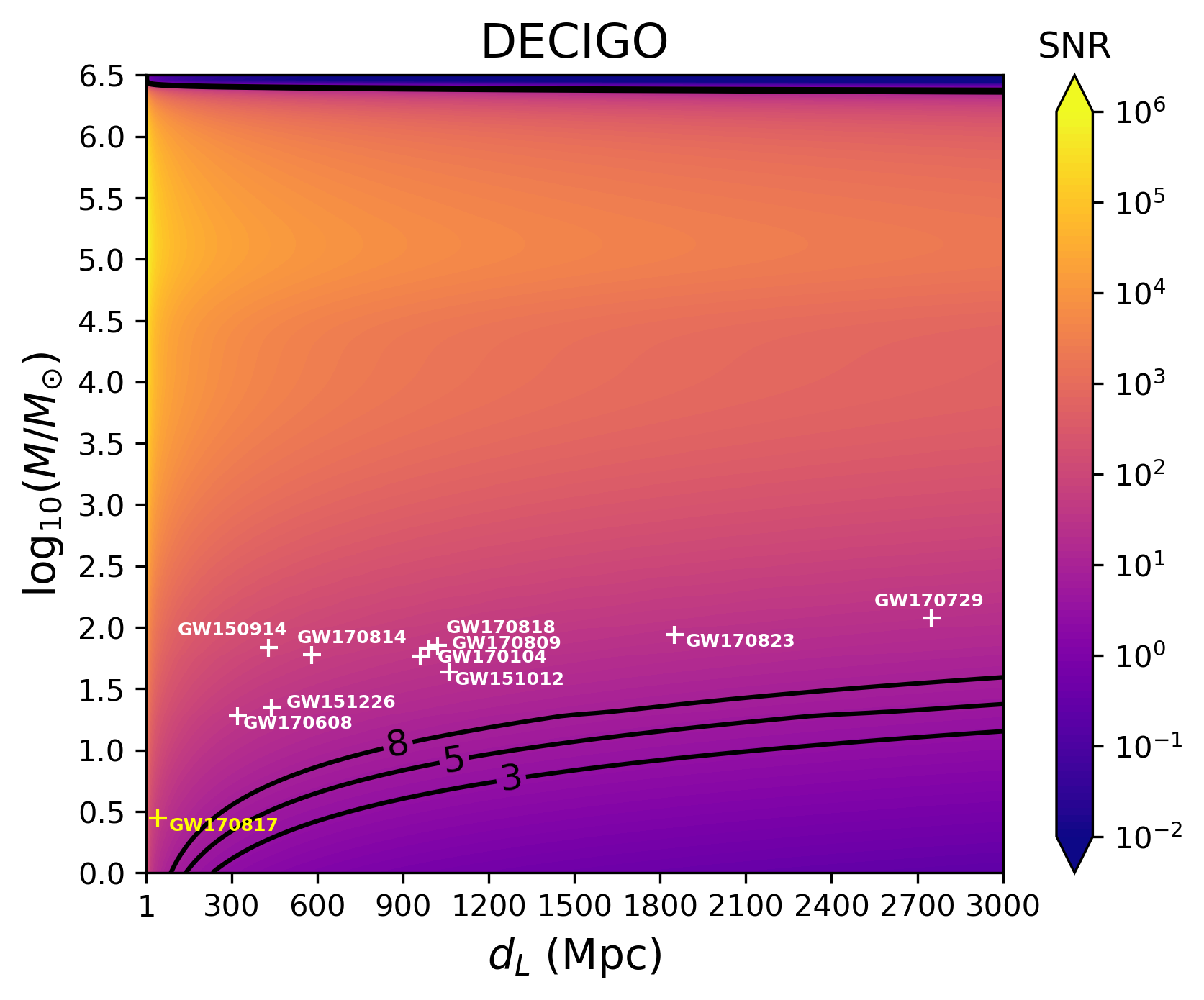} & \includegraphics[width=85mm]{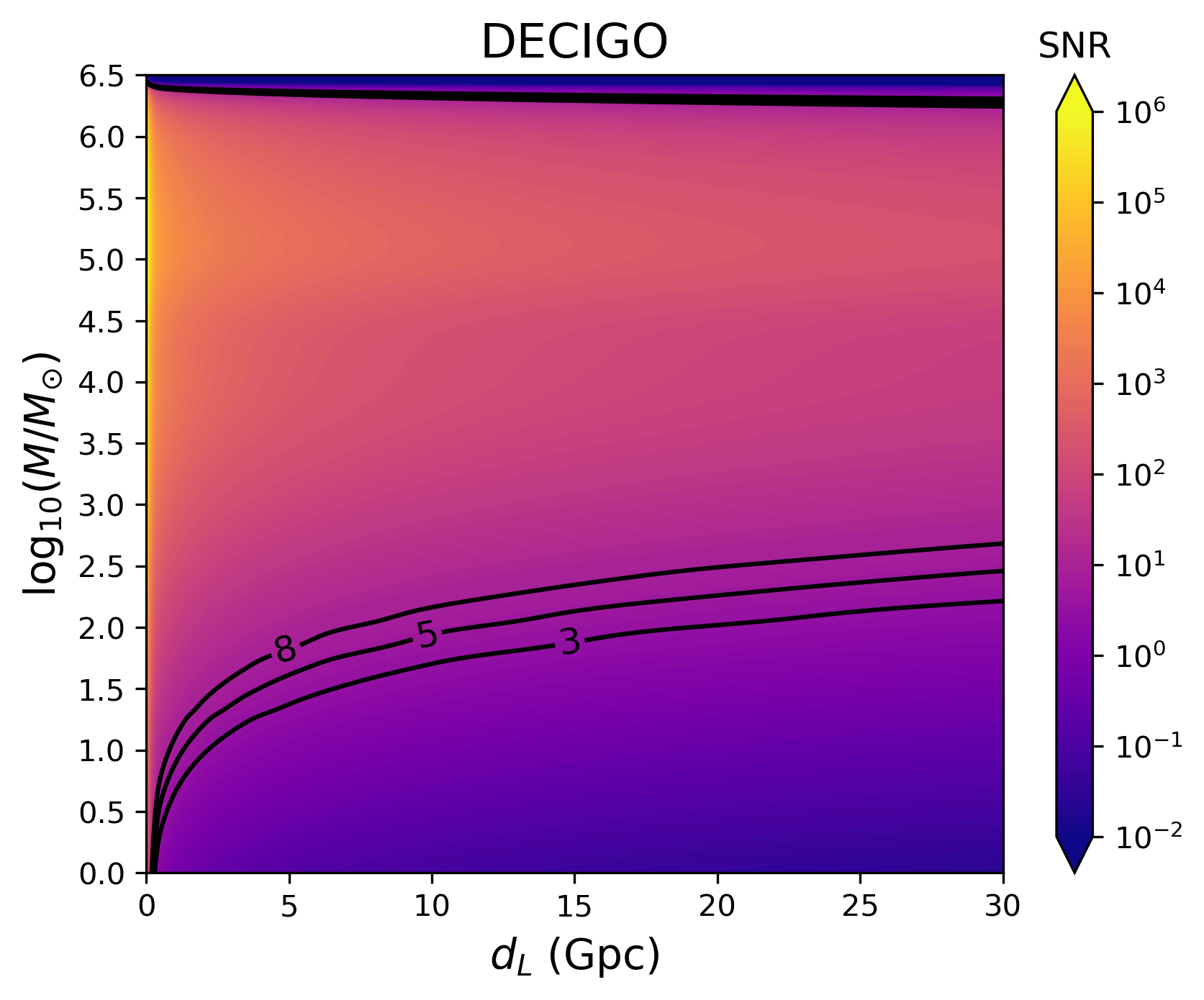} \\
		\includegraphics[width=85mm]{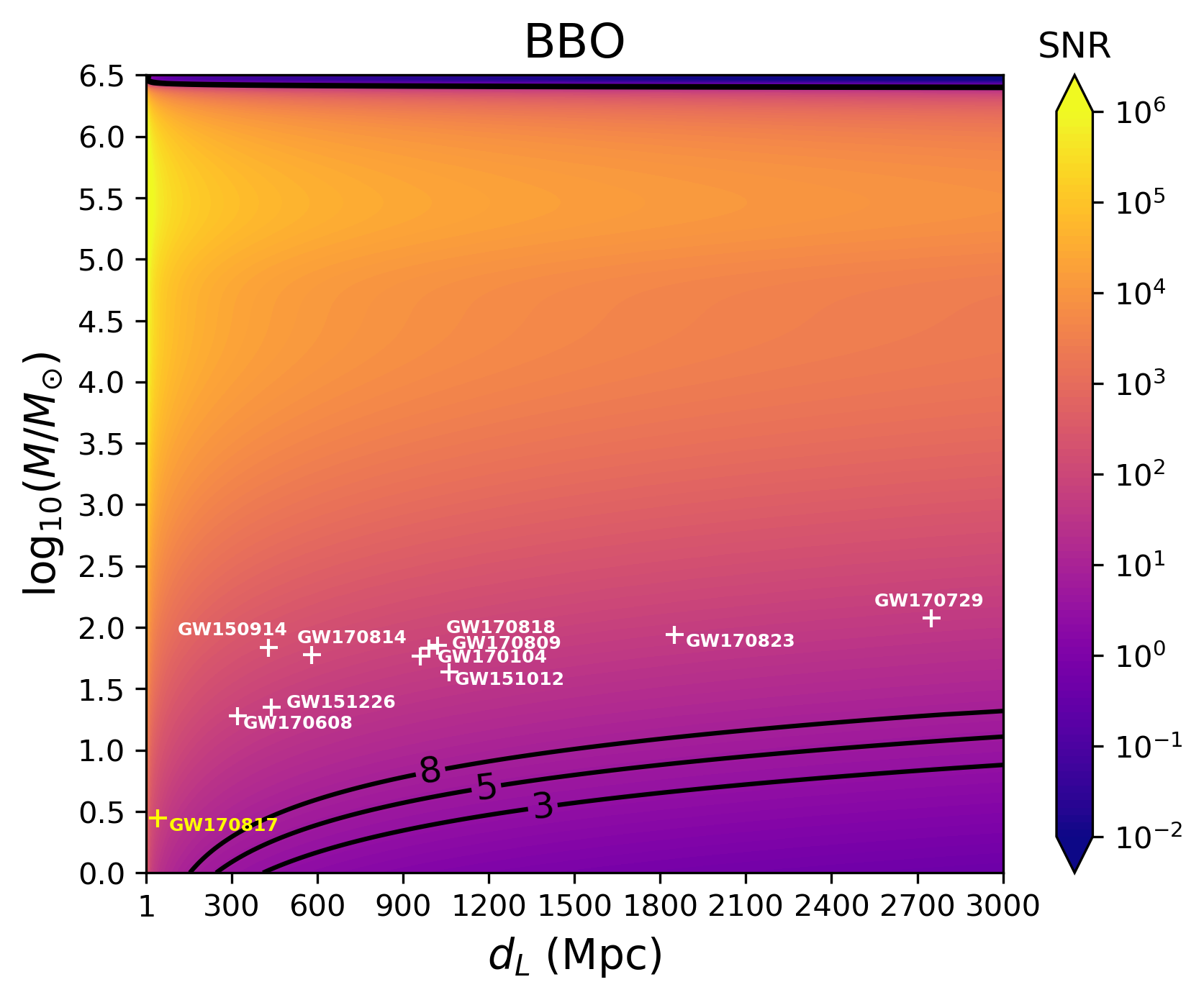} & \includegraphics[width=85mm]{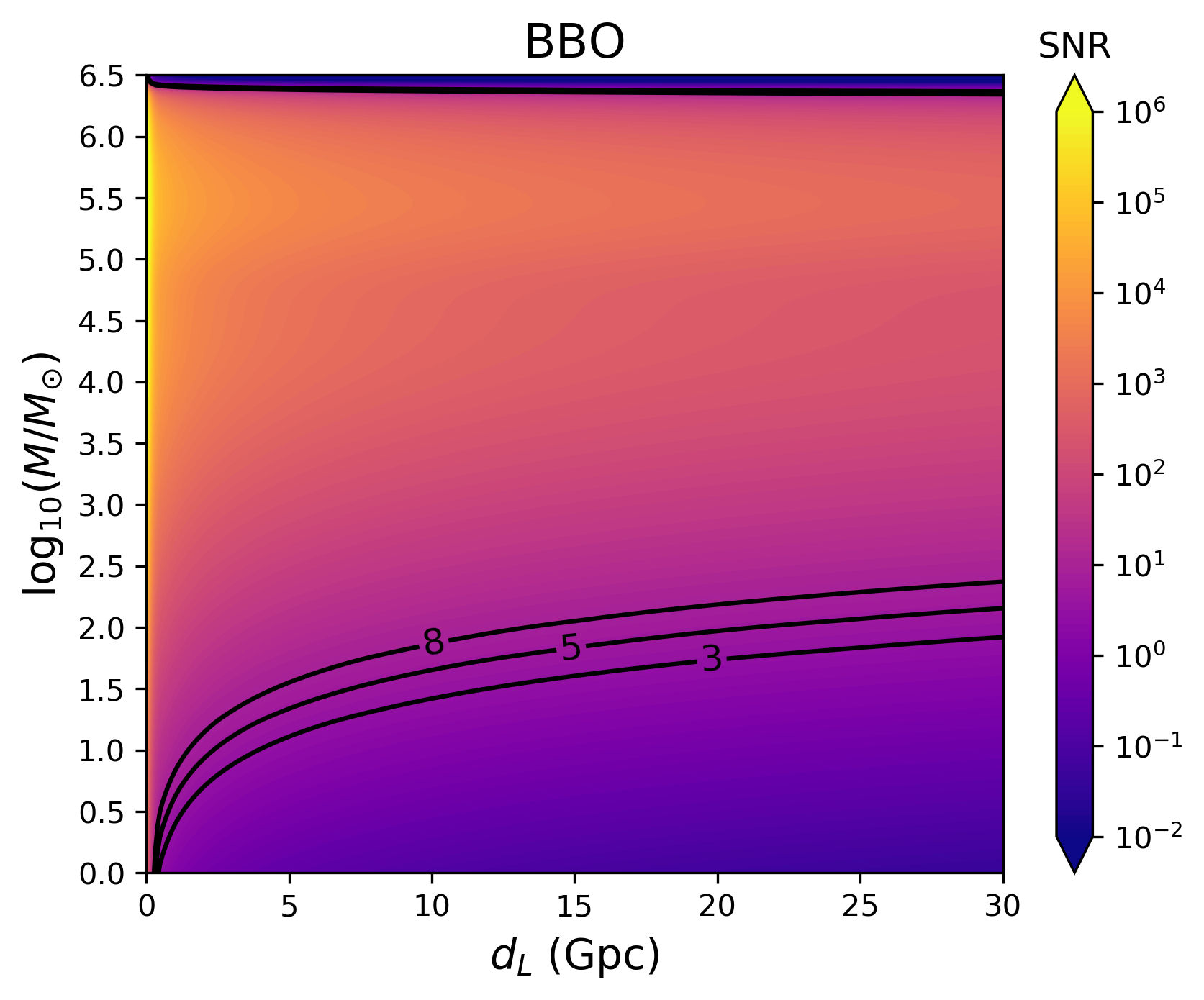}
	\end{tabular}
	\caption{Luminosity distance ($d_L$), total mass (M) parameter space for current and near future ground based detectors. SNR values are shown on a logarithmic color scale for a given $(d_L, M)$ pair. Contours show SNR values of 3, 5, and 8. Events are marked with a ``+" to indicate where they fall on the SNR scale as a BBH, GW150914-like system at optimal orientation. Plots in the left column show SNR values out to 3000 Mpc, and plots on the right column show out to 30 Gpc. Future space based detectors that are sensitive in the dHz regime have excellent sensitivity to the memory from all the released events. The BNS memory might be visible in these detectors. Further study is needed for varying equations of state as in \cite{YangMarty}.}
	\label{spacedet2}
\end{figure}

The reader may have noticed that the memory looks like a Heaviside step function,
\[
	\theta(t) = \begin{cases} 
      0 & t < 0 \\
      1 & t \geq 0
   \end{cases}.
\]
This is one of the functions used in the approximation in \cite{Kennefick}. A step function has a Fourier transform that is proportional to $1/f$. As mass increases in the system, we expect this signal in the frequency domain to increase as it comes into our detector's band and then decrease as it goes out. However, this is not what was found in the ground and space based detectors (Figures \ref{grounddet1}, \ref{grounddet2}, \ref{spacedet1}, \ref{spacedet2}).

The signal in the frequency domain is only well approximated by a step function at low frequencies. The slow step response before merger happens on the radiation reaction timescale and the fast step response during merger happens on the merger timescale. These correspond to features that show up in the frequency domain.

\begin{figure}[H]
	\includegraphics[width=75mm]{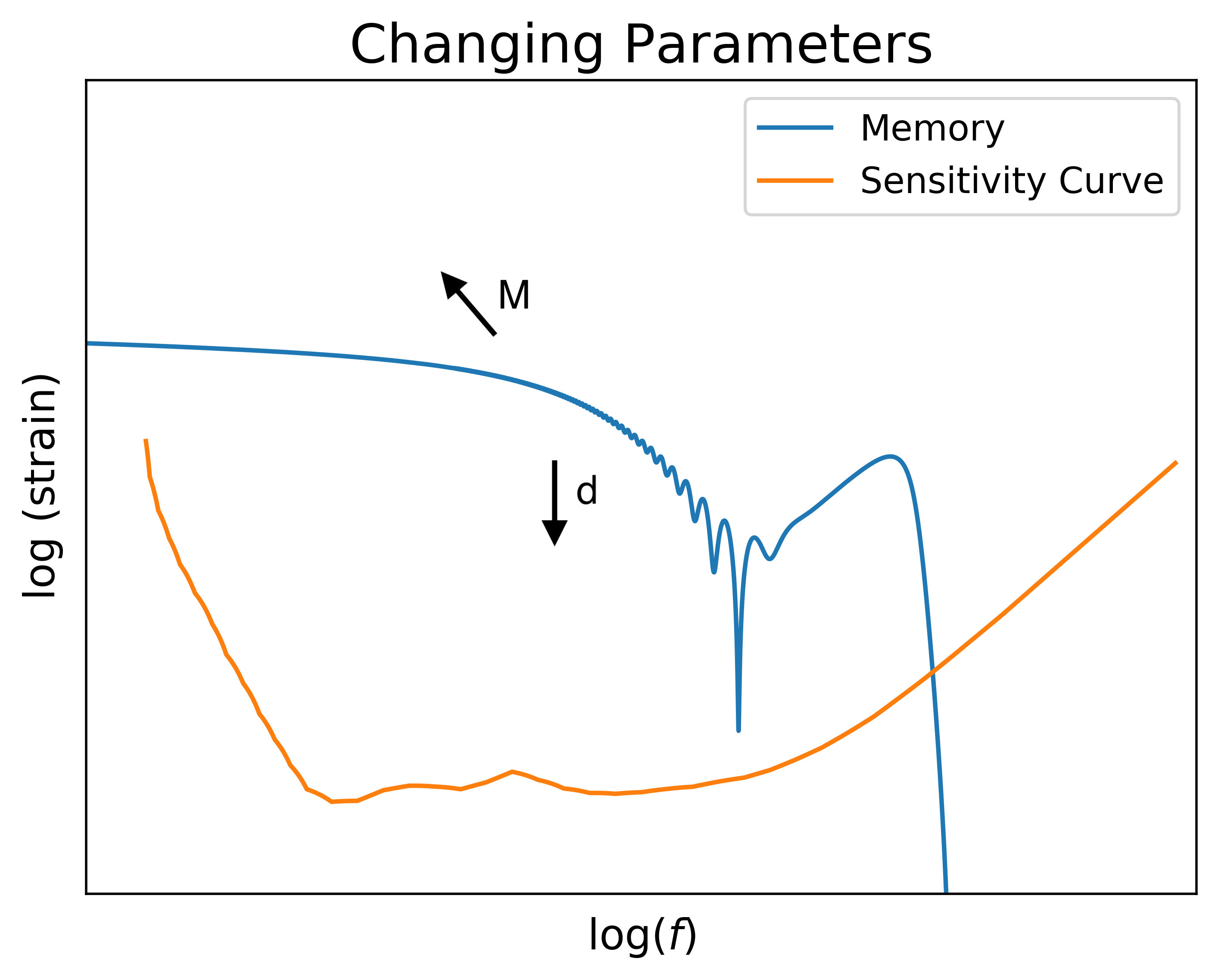}
	\centering
	\caption{Heuristic plot showing the memory signal and a noise curve using the characteristic strain convention. Arrows show the direction the memory moves when changing the total mass, M, and the distance from the source, d.}
	\label{MemNoise}
\end{figure}

Here we use a characteristic strain convention \cite{GWPlot},
\[
	[h_c(f)]^2 = 4 f^2 \left|\tilde{h}(f)\right|^2
\]
and
\[
	[h_n(f)]^2 = f S_n(f).
\]
This is a useful convention for plotting the memory signal over the noise curve, because the area between the two curves on the plot is now proportional to the SNR. However, we must be careful to remember that now a $1/f$ curve on the plot will be a constant. Looking at Figure \ref{MemNoise}, we can see that lower frequencies do behave as $1/f$, but higher frequencies do not! Instead there is a local minimum in the plot and then it falls off completely shortly afterward.

As the mass increases, we find that the curve moves up and to the left. The SNR will increase until the local minimum hits the lowest frequency the detector can see and then it will increase rapidly and fall to zero shortly after. This is the behavior that is seen in Figures \ref{grounddet1}, \ref{grounddet2}, \ref{spacedet1}, \ref{spacedet2}.

\clearpage

\begin{figure}[t]
		\begin{center}
		\Large Pulsar Timing Arrays
		\end{center}
		\begin{tabular}{c | c}
		\textbf{\large Near} & \textbf{\large Far} \\
		\includegraphics[width=85mm]{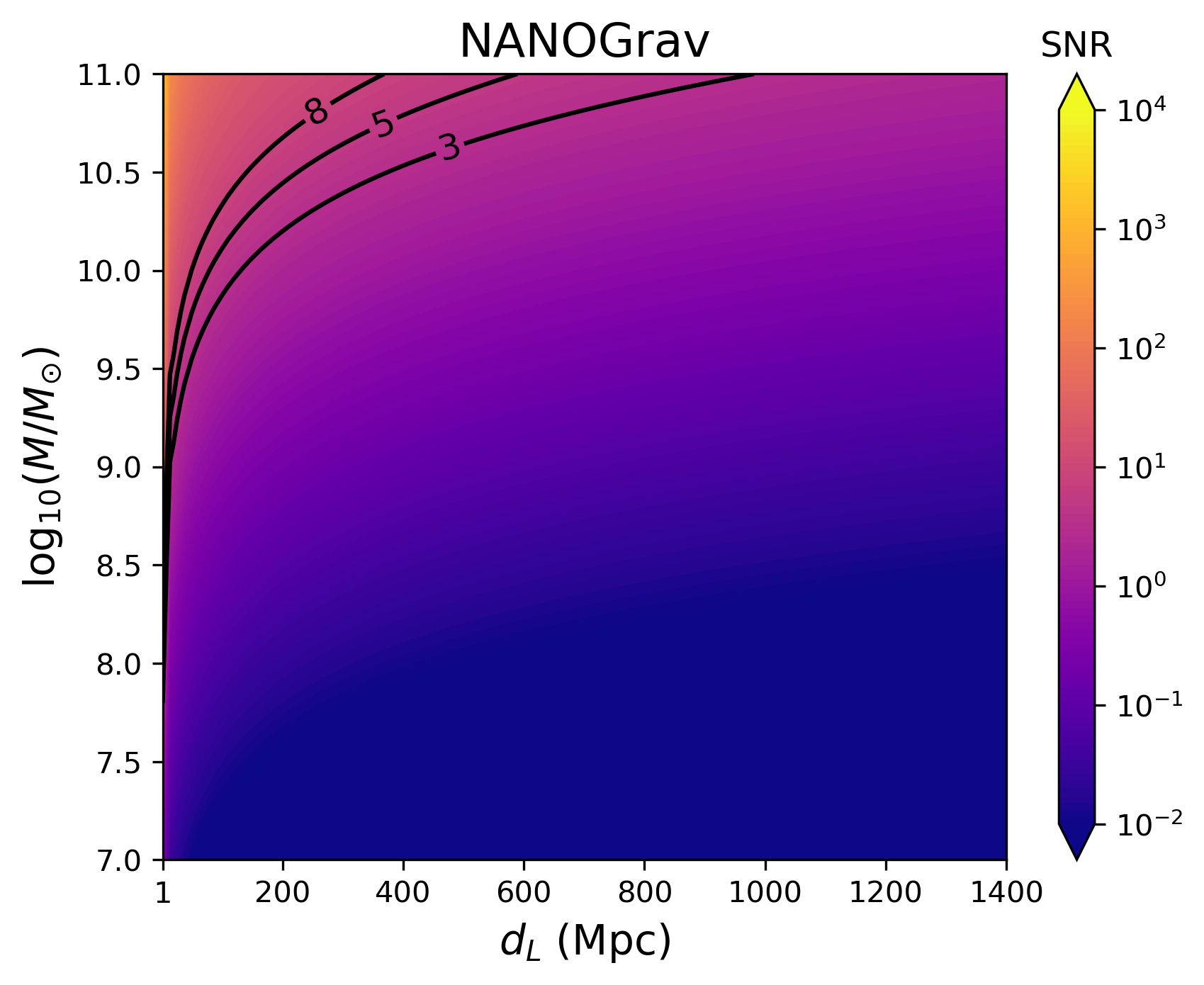} & \includegraphics[width=85mm]{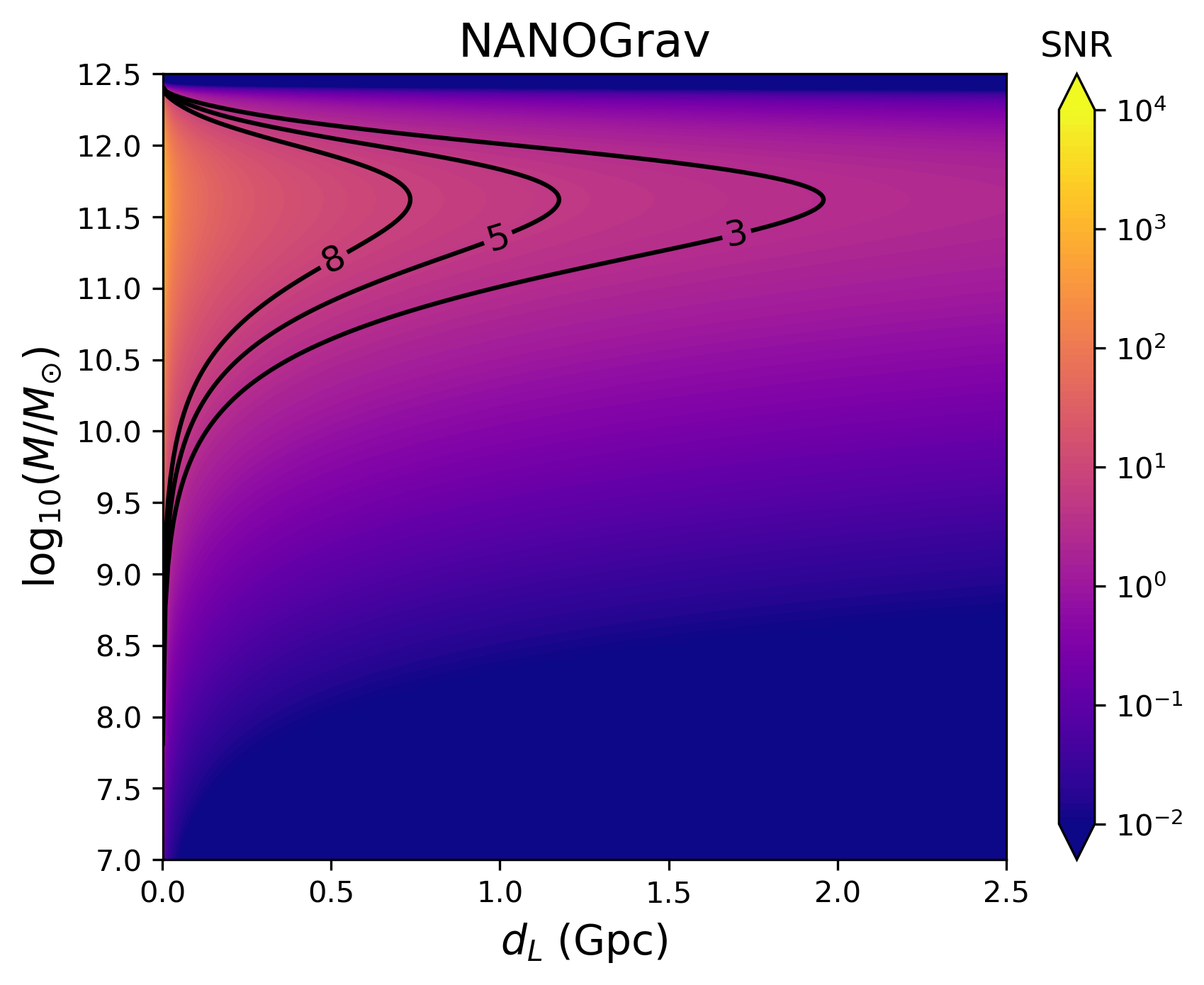} \\
		\includegraphics[width=85mm]{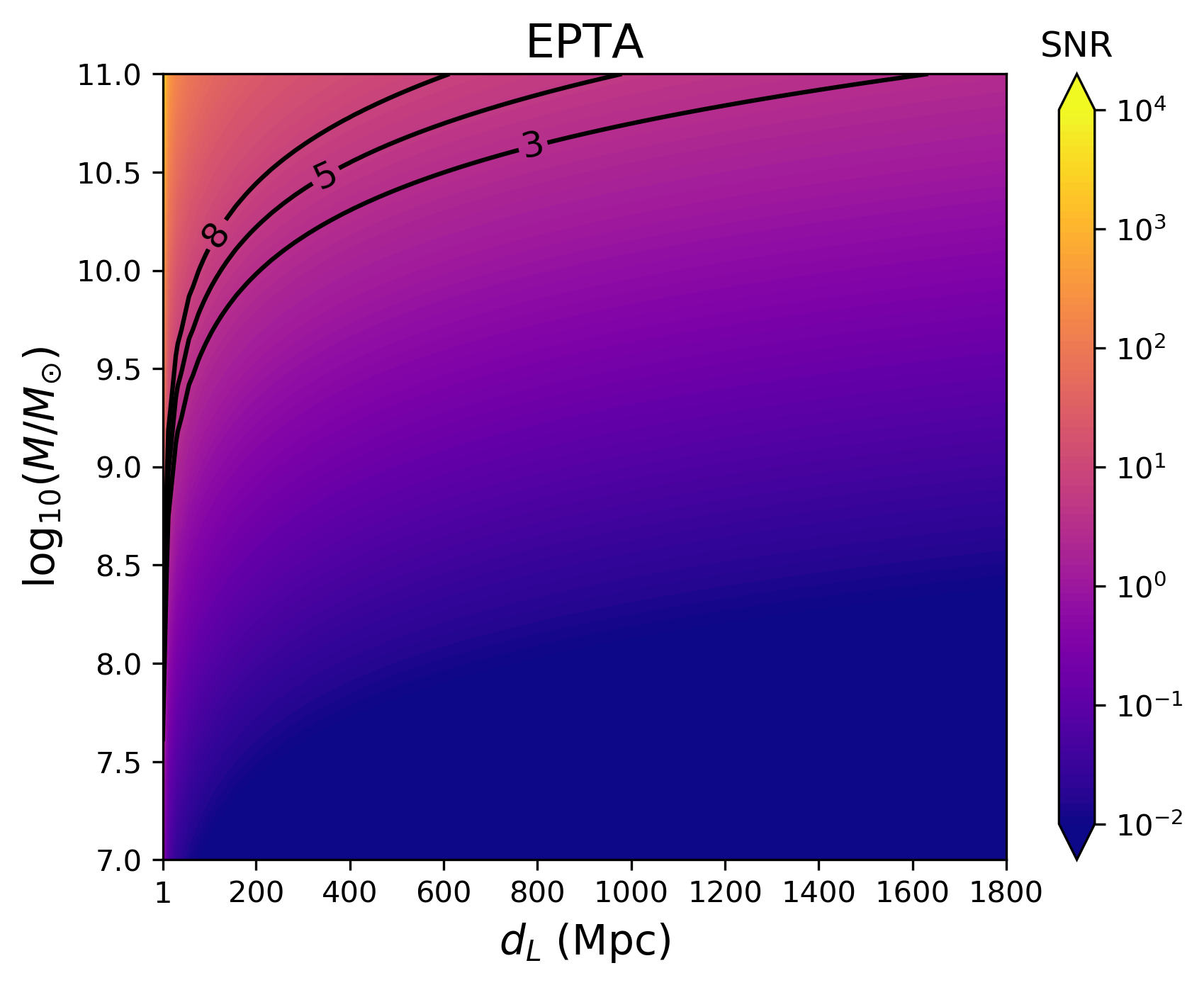} & \includegraphics[width=85mm]{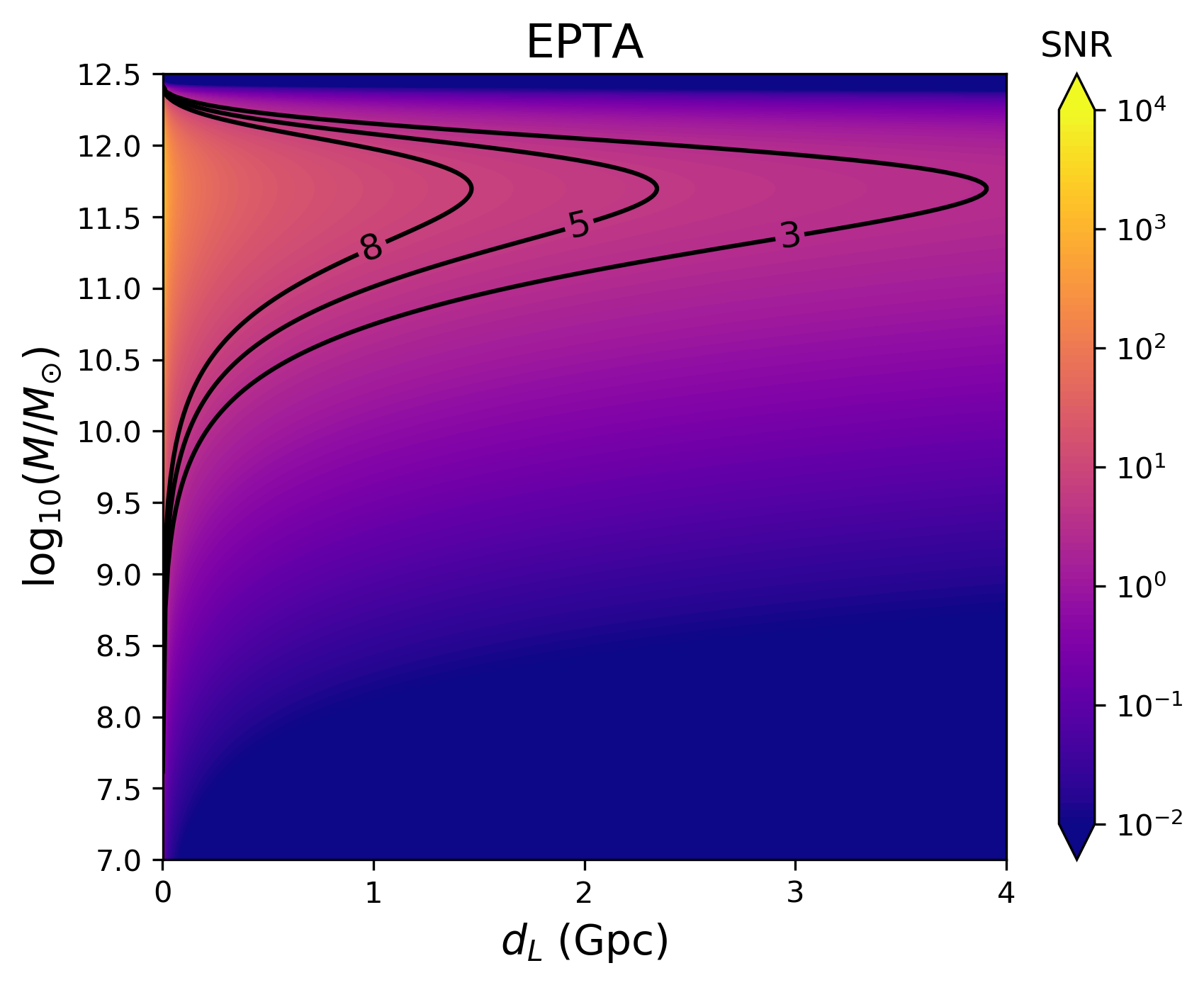}
	\end{tabular}
	\caption{Luminosity distance ($d_L$), total mass (M) parameter space for current and near future ground based detectors. SNR values are shown on a logarithmic color scale for a given $(d_L, M)$ pair. Contours show SNR values of 3, 5, and 8. Events are marked with a ``+" to indicate where they fall on the SNR scale as a BBH, GW150914-like system at optimal orientation. Plots in the left column show SNR values out to 3000 Mpc, and plots on the right column show the edge of the $\rho = 3$ contour. PTAs are sensitive to the memory from the largest SMBH binary systems.}
	\label{PTAdet1}
\end{figure}

\begin{figure}[t]
		\begin{center}
		\Large Pulsar Timing Arrays
		\end{center}
		\begin{tabular}{c | c}
		\textbf{\large Near} & \textbf{\large Far} \\
		\includegraphics[width=85mm]{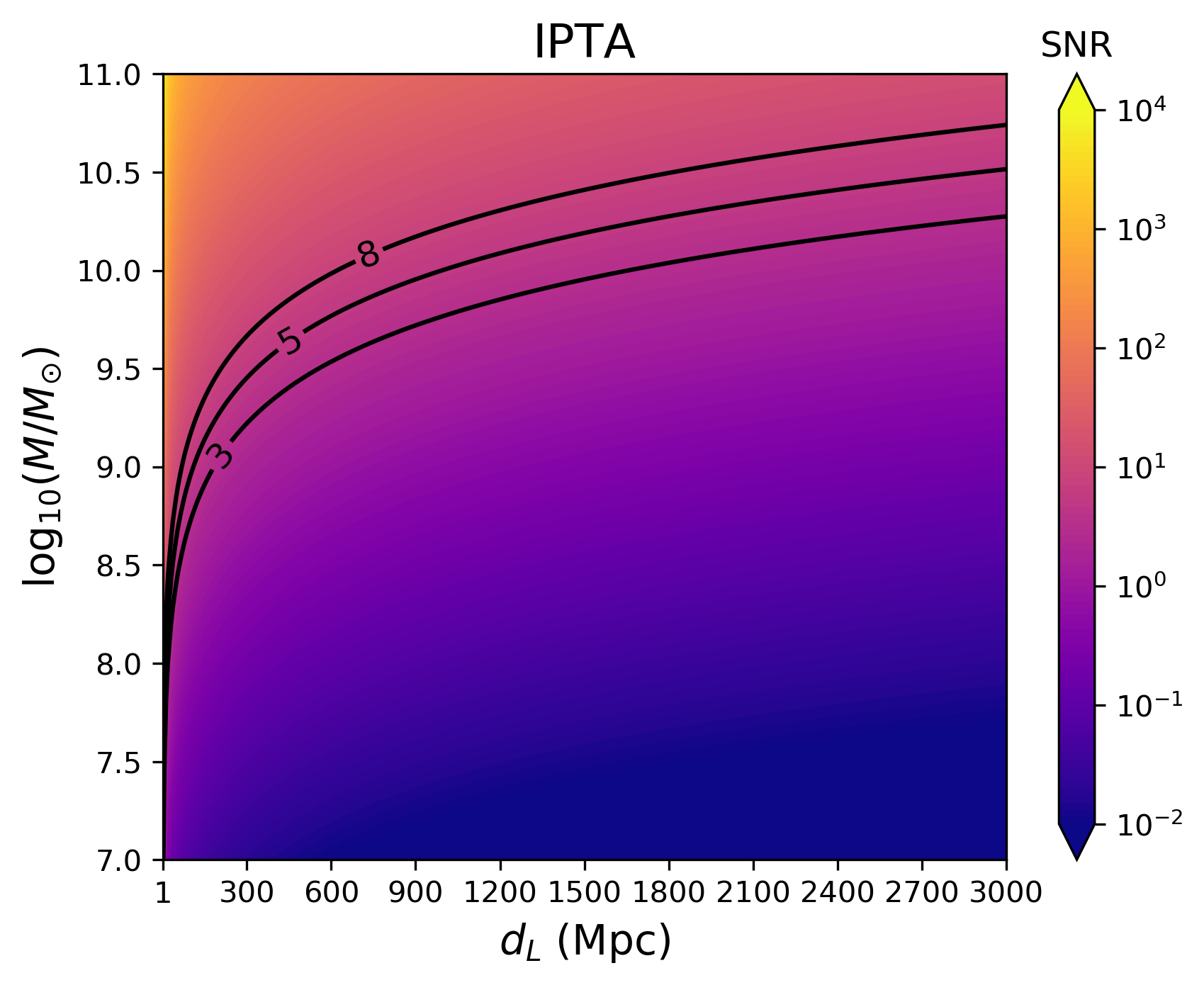} & \includegraphics[width=85mm]{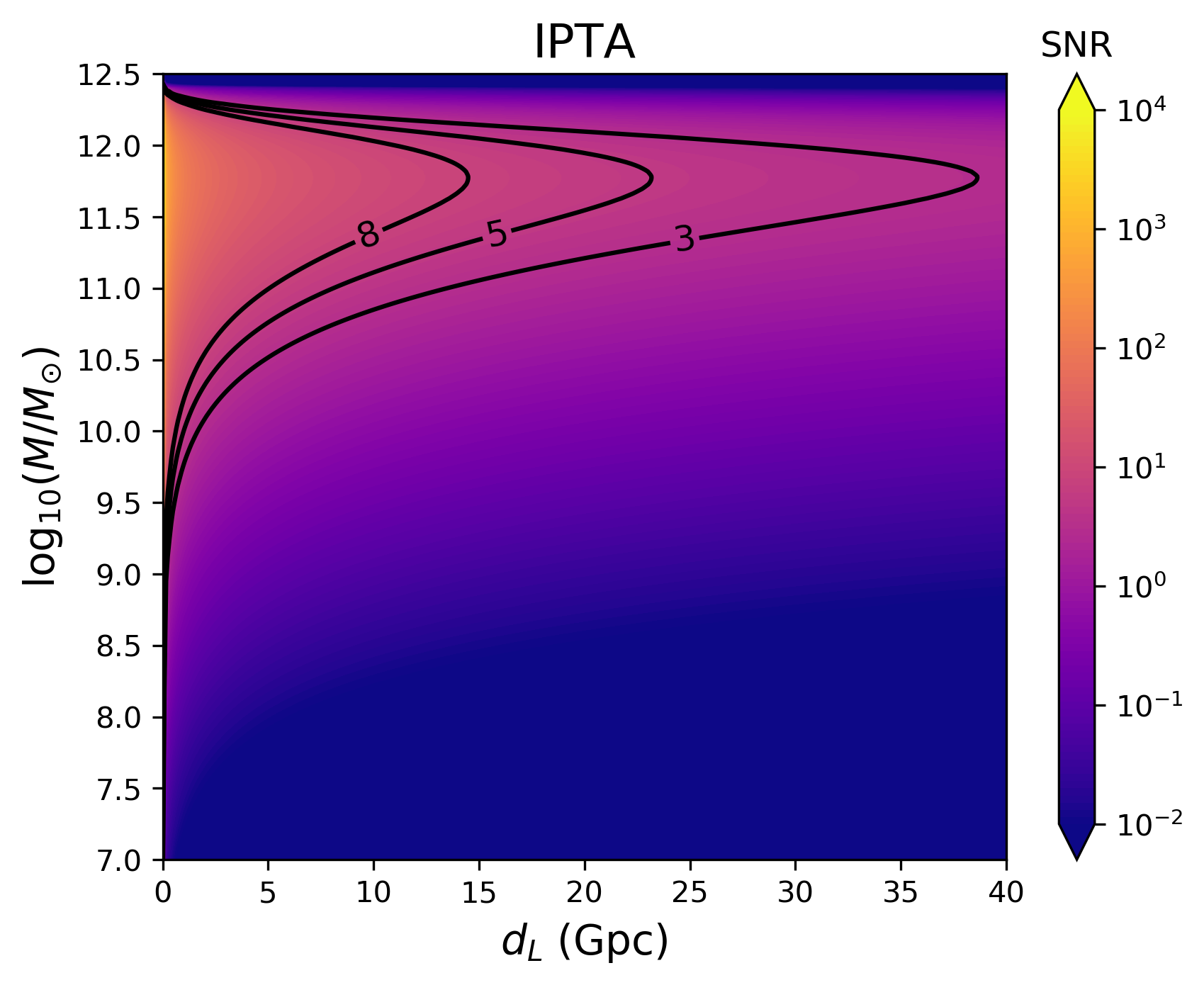} \\
		\includegraphics[width=85mm]{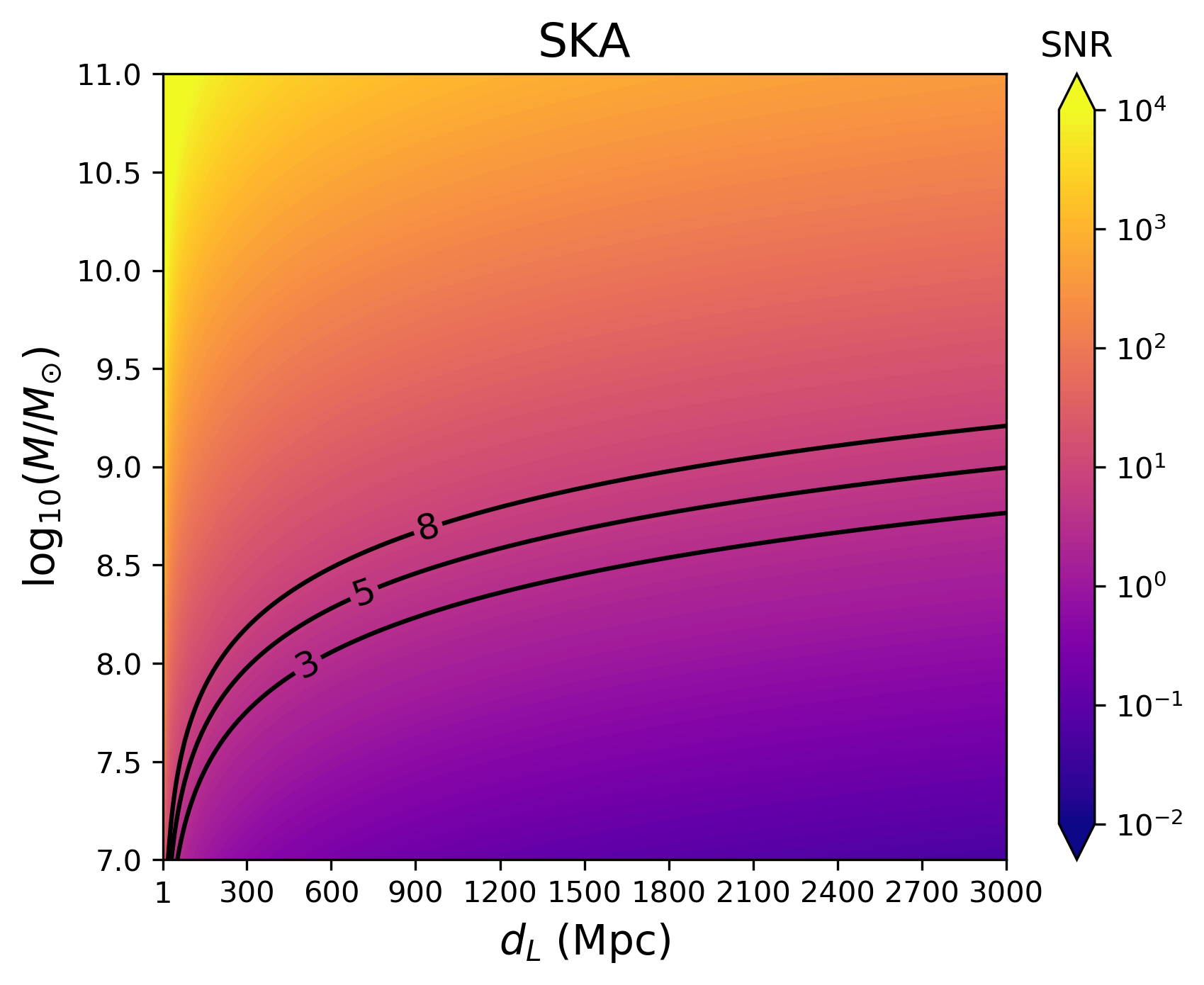} & \includegraphics[width=85mm]{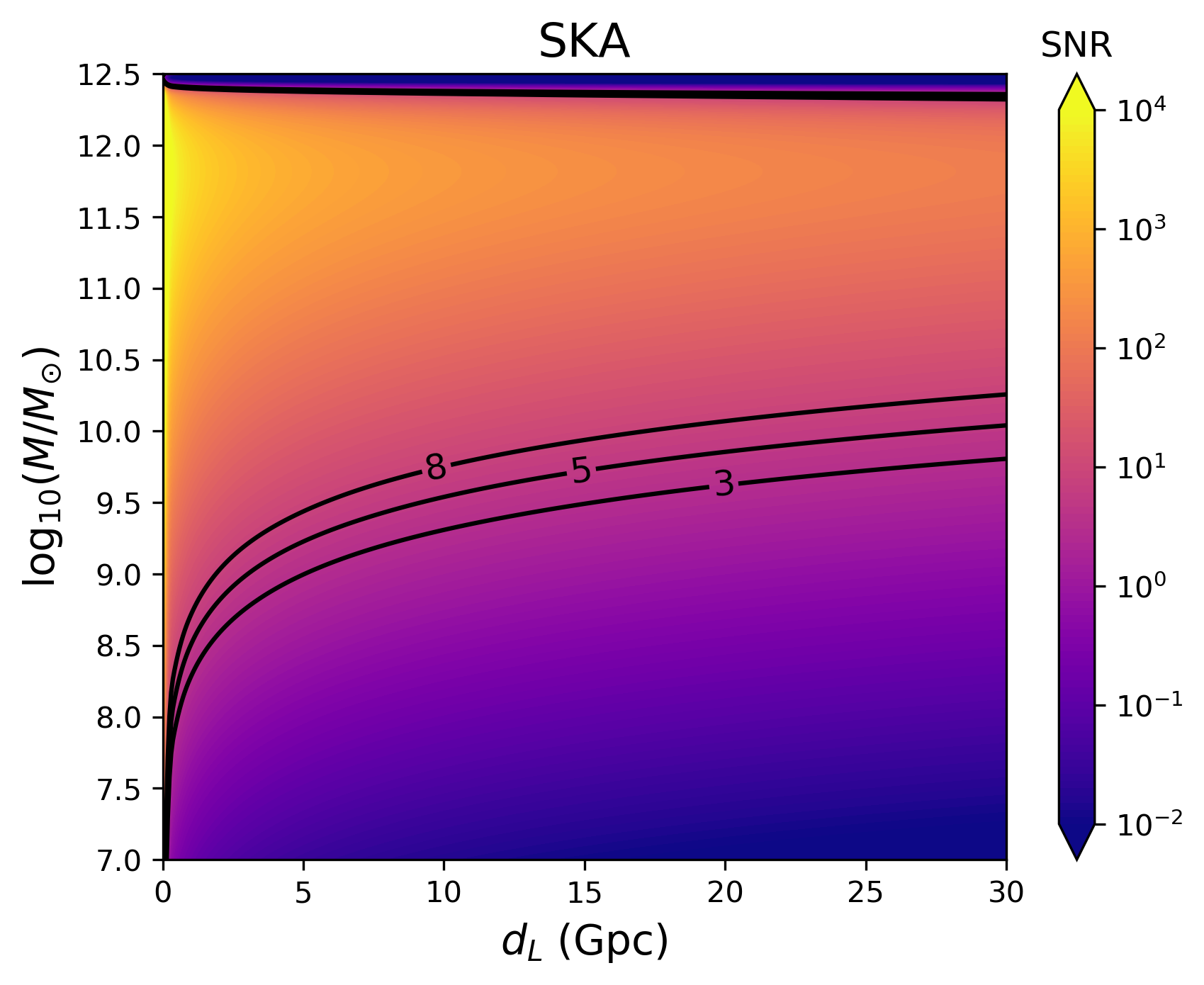}
	\end{tabular}
	\caption{Luminosity distance ($d_L$), total mass (M) parameter space for current and near future ground based detectors. SNR values are shown on a logarithmic color scale for a given $(d_L, M)$ pair. Contours show SNR values of 3, 5, and 8. Events are marked with a ``+" to indicate where they fall on the SNR scale as a BBH, GW150914-like system at optimal orientation. Plots in the left column show SNR values out to 3000 Mpc, and plots on the right column show the edge of the $\rho = 3$ contour or out to 30 Gpc. PTAs are sensitive to the memory from the largest SMBH binary systems.}
	\label{PTAdet2}
\end{figure}

\twocolumngrid

\end{document}